\documentclass[12pt]{iopart}
\usepackage{iopams}  
\usepackage{color}
\usepackage{graphicx}
\usepackage{epsfig}
\usepackage{colordvi}
\usepackage{graphics}
\usepackage{rotating}

\usepackage{dcolumn}
\usepackage{bm}

\newcommand{\bfD}{{\bf D}}%
\newcommand{\bfk}{{\bf k}}%
\newcommand{\bfL}{{\bf L}}%
\newcommand{\bfr}{{\bf r}}%
\newcommand{\bfu}{{\bf u}}%
\newcommand{\bfv}{{\bf v}}%
\newcommand{\bfx}{{\bf x}}%
\newcommand{\bfy}{{\bf y}}%
\newcommand{\bfz}{{\bf z}}%
\newcommand{\bftau}{\boldsymbol{\tau}}%
\newcommand{\bfomega}{\boldsymbol{\omega}}%
\newfont{\tenbfit}{cmbx10}%

\newfont{\tenbbb}{msbm10}%
\newfont{\svnbbb}{msbm8}%
\newcommand{\half}{{\textstyle{\frac{1}{2}}}}

\newcommand{\Grad}{\hbox{\rm grad}\mskip2mu}
\newcommand{\Div}{\hbox{\rm div}\mskip2mu}
\newcommand{\Curl}{\hbox{\rm curl}\mskip2mu}
\newcommand{\trans}{\mskip-2mu\scriptscriptstyle\top\mskip-2mu}

\newcommand{\lj}{\mbox{$[\kern-0.1478125em[$}}
\newcommand{\rj}{\mbox{$]\kern-0.1478125em]$}}
\newcommand{\la}{\mbox{$\langle\kern-0.2325em\langle$}}
\newcommand{\ra}{\mbox{$\rangle\kern-0.2325em\rangle$}}
\newcommand{\Blj}{\mbox{$\Big[\kern-0.275em\Big[$}}
\newcommand{\Brj}{\mbox{$\Big]\kern-0.275em\Big]$}}
\newcommand{\Bla}{\mbox{$\Big\langle\kern-0.425em\Big\langle$}}
\newcommand{\Bra}{\mbox{$\Big\rangle\kern-0.425em\Big\rangle$}}

\newcommand{\zed}{{\bf 0}}

\newcommand{\C}{\Curl}

\newcommand{\ddt}{\frac{\rm{d}}{{\rm{d}}t}}
\newcommand{\dt}[1]{\frac{{\rm{d}}#1}{{\rm{d}}t}}

\newcommand{\dv}{\,{\rm{d}}v}

\DeclareSymbolFont{AMSa}{U}{msa}{m}{n}
\DeclareSymbolFont{AMSb}{U}{msb}{m}{n}
\DeclareSymbolFontAlphabet{\mathbb}{AMSb}%

\begin{document}


\title[Statistics of the Navier--Stokes-alpha-beta regularization model]{Statistics of the Navier--Stokes-alpha-beta regularization model for fluid turbulence}

\author{ Denis F.\ Hinz}
\address{Department of Mechanical Engineering, McGill University,\\  Montr\'eal, QC H3A 2K6, Canada}
\ead{denis.hinz@mail.mcgill.ca}

\author{ Tae-Yeon Kim}
\address{Department of Mechanical Engineering, University of Washington,  Seattle, WA 98195, USA}
\ead{tykimsay@gmail.com}

\author{Eliot Fried }
\address{Department of Mechanical Engineering, University of Washington,  Seattle, WA 98195, USA}
\ead{mechanicist@gmail.com}



\begin{abstract}
We explore one-point and two-point statistics of the Navier--Stokes-$\alpha\beta$ regularization model at moderate Reynolds number ($Re \approx 200$) in homogeneous isotropic turbulence. The results are compared to the limit cases of the Navier--Stokes-$\alpha$ model and the Navier--Stokes-$\alpha\beta$ model without subgrid-scale stress, as well as with high resolution direct numerical simulation. After reviewing spectra of different energy norms of the Navier--Stokes-$\alpha\beta$ model, the Navier--Stokes-$\alpha$ model, and Navier--Stokes-$\alpha\beta$ model without subrid-scale stress, we present probability density functions and normalized probability density functions of the filtered and unfiltered velocity increments along with longitudinal velocity structure functions of the regularization models and direct numerical simulation results. We highlight differences in the statistical properties of the unfiltered and filtered velocity fields entering the governing equations of the Navier--Stokes-$\alpha$ and Navier--Stokes-$\alpha\beta$ models and discuss the usability of both velocity fields for realistic flow predictions. The influence of the modified viscous term in the Navier--Stokes-$\alpha\beta$ model is studied through comparison to the case where the underlying subgrid-scale stress tensor is neglected. The filtered velocity field is found to have physically more viable probability density functions and structure functions for the approximation of direct numerical simulation results, whereas the unfiltered velocity field is found to have flatness factors close to direct numerical simulation results.
\end{abstract}
\maketitle

\section{Introduction}
In principle, the accurate prediction of turbulent flows implies the resolution of the motion of the fluid on all spatial scales ranging from the large ones determined by the geometry of the flow domain to the smallest ones, at which the kinetic energy of fluid particles is transformed into heat by molecular dissipation. Even with access to state-of-the-art supercomputers, direct numerical simulations (DNS) of turbulent flows, meaning the solution of the Navier-Stokes (NS) equations at very high spatial and temporal resolution, are intractable, except for particularly simple flow configurations. While DNS is widely used as a research tool to perform numerical experiments and explore the physical properties of turbulence, practical flow predictions, e.g.\ in realistic engineering or geophysical flows, are founded on turbulence models, which attempt to include the physical effects of turbulence at significantly lower computational costs than DNS. Turbulence models aim at predicting relevant large-scale flow features without resolving the smallest scales of the fluctuating turbulent velocity field. The interaction between the large scales and the small scales is parametrized in turbulence models. 

The class of regularized turbulence models stems originally from the idea of performing statistical averaging at the level of the action principle using the Hamiltonian formalism and obtaining closure at the variational level with Taylor's ``frozen-in turbulence" hypothesis, as discussed by Holm~\cite{Holm2005}. The Navier--Stokes-$\alpha$ (NS-$\alpha$) model (also called the viscous Camassa--Holm equations or Lagrangian averaged Navier--Stokes (LANS-$\alpha$) model) was first obtained by Chen et al.\ \cite{Chen1998, Chen1999, Chen1999a} by adding a viscous term to the inviscid Camassa--Holm equations \cite{Holm1998a, Holm1998}, (also called the Lagrangian averaged Euler (LAE) equations or Euler-$\alpha$ equations). For a statistically homogeneous and isotropic flow with constant density $\rho$ and constant dynamic viscosity $\mu$, the NS-$\alpha$ model constitutes a system
\begin{equation}\label{eq:NSalpha}
\left.
\begin{array}{c}
\displaystyle
\rho \Big(\frac{\partial \bfv }{\partial t} + (\Grad \bfv) \bfu + (\Grad \bfu)^{\trans} \bfv \Big)
=-\Grad \varpi + \mu (1-\alpha^2\triangle) \triangle \bfu,
\cr\noalign{\vskip8pt}
\bfv  = ( 1 - \alpha^2 \triangle ) \bfu, \qquad  \Div \bfu = 0, 
\end{array}
\!\!
\right\}
\end{equation}
for filtered and unfiltered velocities $\bfu$ and $\bfv$ related, as indicated in \eref{eq:NSalpha}$_2$, through the modified Helmholtz operator $( 1 - \alpha^2 \triangle )$, and a filtered pressure-like variable $\varpi= p- \half \rho(|\bfu|^2 +\alpha^2 |\bfD|^2)$, with $\bfD = \half (\Grad \bfu + (\Grad \bfu) ^{\trans})$ the filtered stretching tensor. In the limit $\alpha \rightarrow 0$, the NS equations are recovered from~\eref{eq:NSalpha}. Various interpretations of the filtered and unfiltered velocities have been provided by Holm et al.~\cite{Holm1999a}. In view of~\eref{eq:NSalpha}$_2$, the velocity $\bfu$ is smoothed by applying the inverse ${\cal H}_\alpha^{-1}$ of the modified Helmholtz operator 
\begin{equation}\label{eq:Helmholtz01}
{\cal H}_\alpha=1-\alpha^2\triangle
\end{equation}
to the unfiltered velocity $\bfv$, and thus contains less information than $\bfv$. The constant filter parameter $\alpha$ has dimensions of length, and heuristically, represents the characteristic linear dimension of the smallest eddy that the model can resolve. Numerical simulations indicate that the NS-$\alpha$ model captures the properties of flows for eddy scales greater than $\alpha$ (see, for example, Chen et al.~\cite{Chen1999b}).

\section{The Navier--Stokes-$\alpha\beta$ model}

Based on a framework for fluid theories with higher-order gradient dependencies, Fried \& Gurtin \cite{Fried2007, Fried2010, Fried2008} proposed a slight generalization of the NS-$\alpha$ model, which they call Navier--Stokes-$\alpha\beta$ (NS-$\alpha\beta$) model. For a statistically homogeneous and isotropic turbulent flow of an incompressible fluid with constant mass density $\rho$ and constant dynamic viscosity $\mu$, the NS-$\alpha\beta$ model constitutes a system
\begin{equation}\label{eq:NSalphabeta}
\left.
\begin{array}{c}
\displaystyle
\rho \Big(\frac{\partial \bfv }{\partial t} + (\Grad \bfv) \bfu + (\Grad \bfu)^{\trans} \bfv \Big)
=-\Grad \varpi + \mu (1-\beta^2\triangle) \triangle \bfu,
\cr\noalign{\vskip8pt}
\bfv  = ( 1 - \alpha^2 \triangle ) \bfu, \qquad  \Div \bfu = 0, 
\end{array}
\!\!
\right\}
\end{equation}
very similar to the system of the NS-$\alpha$ equations~\eref{eq:NSalpha}. Whereas the parameter $\alpha > 0$ is the Helmholtz
filtering radius, as in the NS-$\alpha$ model, and is meant to be representative of eddy scales in the inertial range, the parameter $ \beta > 0$ is meant to be representative
of eddy scales in the dissipation range; Fried \& Gurtin \cite{Fried2007,Fried2008, Fried2010} thus expect that $\beta < \alpha$.
Importantly, setting $\beta=\alpha$ in~\eref{eq:NSalphabeta}$_1$ yields the NS-$\alpha$ equations~\eref{eq:NSalpha} and the NS equations are recovered in the limit $\alpha,\beta \rightarrow 0$. Notice that the only difference between the NS-$\alpha$ and NS-$\alpha\beta$ models is the modification of the viscous term as shown in~\eref{eq:NSalphabeta}$_1$ and~\eref{eq:NSalpha}$_1$. 

In view of the identity
\begin{equation}
\Grad (\bfu\cdot\bfv) = (\Grad\bfv)\bfu+(\Grad\bfu)^{\trans}\bfv+\bfu\times\C\bfv
\end{equation}
and the definition $\nu = \mu /\rho$ of the kinematic viscosity, the flow equation~\eref{eq:NSalphabeta}$_1$ can be written in the alternative form
\begin{equation}\label{eq:NSabRot}
\frac{\partial \bfv }{\partial t} - \bfu \times \C \bfv
= -\Grad \frac{\varpi}{\rho} + \nu (1-\beta^2\triangle) \triangle \bfu.
\end{equation}
Dotting each term of~\eref{eq:NSabRot} with $\bfu$ and integrating over the flow domain $R$ with periodic boundary conditions imposed on $\partial R$ yields an energy balance
\begin{equation}\label{eq:NSabebalance}
\ddt \int\limits_R \half ( |\bfu|^2+\alpha^2|\bfomega|^2) \dv 
=
- \int\limits_R \nu(|\bfomega|^2+\beta^2|\Grad \bfomega|^2 ) \dv,
\end{equation}
where $\bfomega = \Curl \bfu$ is the filtered vorticity. With reference to~\eref{eq:NSabebalance}, the kinetic energy $E$ and the dissipation rate $\epsilon$ for the model, both measured per unit mass, are given by
\begin{equation}\label{eq:NSabkenergy}
\left.
\begin{array}{c}
\displaystyle E =\int\limits_R \half ( |\bfu|^2+\alpha^2|\bfomega|^2) \dv,\\
\displaystyle \epsilon =\int\limits_R \nu(|\bfomega|^2+\beta^2|\Grad \bfomega|^2 ) \dv.\\
\end{array}
\right\}
\end{equation}
Notice that the kinetic energy~\eref{eq:NSabkenergy}$_1$ of the NS-$\alpha\beta$ model is identical to that of the NS-$\alpha$ model. Setting $\beta=\alpha$ in ~\eref{eq:NSabkenergy}$_2$ recovers the dissipation rate of the NS-$\alpha$ model. Thus, choosing $\beta<\alpha$ reduces the dissipation of the kinetic energy relative to the dissipation rate of the NS-$\alpha$ model.

To derive an alternative LES form of the NS-$\alpha\beta$ model, we first record the identities
\begin{equation}\label{eq:id01}
(\Grad \bfu)^{\trans} \bfv   =  \half \Grad(|\bfu|^2+\alpha^2 |\bfL|^2)-\alpha^2 \Div(\bfL^{\trans} \bfL),
\end{equation}
and
\begin{equation}\label{eq:id02}
\dot{\overline{{\cal H}_\alpha\bfu}} = {\cal H}_\alpha\dot{\bfu} +\alpha^2 \Div(\bfL\bfL + \bfL\bfL^{\trans}),
\end{equation}
where a superposed dot denotes material time differentiation following $\bfu$, $\bfL = \Grad \bfu$, and the constraint $\Div \bfu =0$ has been used. In view of~\eref{eq:Helmholtz01}, \eref{eq:NSalphabeta}$_2$, \eref{eq:id01}, and~\eref{eq:id02}, we may rewrite~\eref{eq:NSalphabeta}$_1$ as
\begin{equation}\label{eq:les01}
\rho {\cal H}_\alpha\dot{\bfu} = -\Grad  P  
- \rho\alpha^2\Div (\bfL \bfL^{\trans} + \bfL \bfL - \bfL^{\trans}\bfL  ) + \mu (1- \beta^2\triangle) \triangle \bfu,
\end{equation}
where $P = \varpi + \half \rho (|\bfu|^2 + \alpha^2 |\bfL|^2 )$ is a pressure-like variable. Next, by~\eref{eq:Helmholtz01},~\eref{eq:les01} is equivalent to
\begin{equation}\label{eq:les02}
\fl \rho {\cal H}_\alpha\dot{\bfu} = -\Grad  P  - \rho\alpha^2\Div (\bfL \bfL^{\trans} + \bfL \bfL - \bfL^{\trans}\bfL  ) 
+ \mu {\cal H}_\alpha\triangle \bfu+ \mu (\alpha^2- \beta^2) \triangle \triangle \bfu,
\end{equation}
which, on applying the inverse modified Helmholtz operator ${\cal H}_\alpha^{-1}$, yields the LES form 
\begin{equation}\label{eq:les03}
\fl \rho \dot{\bfu} = -\Grad  \overline{P}  -  \rho\alpha^2{\cal H}_\alpha^{-1} \Div (\bfL \bfL^{\trans} + \bfL \bfL - \bfL^{\trans}\bfL  ) 
 +\mu \triangle \bfu+ \mu (\alpha^2- \beta^2) {\cal H}_\alpha^{-1} \triangle \triangle \bfu
\end{equation}
of the NS-$\alpha\beta$ equations. Notice that~\eref{eq:les03} involves only the filtered velocity $\bfu$ and a filtered pressure-like variable $\overline{P}$ .
Equivalently, on defining the subgrid-scale (SGS) stress tensor $\bftau_\alpha=\rho\alpha^2{\cal H}_\alpha^{-1}(\bfL\bfL^{\trans}+\bfL\bfL-\bfL^{\trans}\bfL)$, the LES form  of the NS-$\alpha\beta$ model constitutes a system
\begin{equation}\label{eq:NSalphaC2}
\left.
\begin{array}{l}
\rho \dot{\bfu} = -\Grad \overline P  - \Div\bftau_\alpha
+\mu \triangle \bfu + \mu (\alpha^2 - \beta^2) {\cal H}_\alpha^{-1} \triangle \triangle \bfu, 
\cr\noalign{\vskip8pt}
\bftau_\alpha=  \rho \alpha^2{\cal H}_\alpha^{-1}  (\bfL \bfL^{\trans} + \bfL \bfL - \bfL^{\trans}\bfL ), 
\cr\noalign{\vskip8pt}
\Div \bfu = 0. 
\end{array}
\!\!
\right\}
\end{equation}
Importantly, the systems~\eref{eq:NSalphaC2} and~\eref{eq:NSalphabeta} are equivalent. From~\eref{eq:NSalphaC2} it is evident that the NS-$\alpha\beta$ model has the same SGS stress tensor as the NS-$\alpha$ model. For a discussion of the SGS stress terms of the NS-$\alpha$ model and related models of the same family see, for example, Geurts et al.~\cite{Geurts2002}. In the LES form, the difference between the NS-$\alpha\beta$ model and the NS-$\alpha$ model manifests itself by an additional viscous term $\mu (\alpha^2 - \beta^2) {\cal H}_\alpha^{-1} \triangle \triangle \bfu$, which vanishes for the limit case $\beta=\alpha$. This additional modeling term is different in nature from the terms involved in the SGS stress in the sense that it carries as a factor the viscosity coefficient $\mu$. Foias et al.~\cite{Foias2001a, Foias2002} discuss the LES form of the NS-$\alpha$ model analytically, and Geurts et al.~\cite{Geurts2002} solve the LES form of the NS-$\alpha$ model along with other models of the same family numerically.

The equations in~\eref{eq:NSalphabeta} are non-dimensionalized by introducing a characteristic length scale $L$, a characteristic velocity scale $U$, and a characteristic time scale $L/U$. For convenience, we maintain the symbols used for the original dimensional variables. As a result, the dimensionless viscosity $\nu$ represents the inverse Reynolds-number of the flow. The dimensionless length scales $\alpha$ and $\beta$ arise simply by dividing their dimensional counterparts by $L$. For numerical simulations of homogeneous, isotropic turbulence in a periodic box it is convenient to choose the dimensionless side length of the box to be $2\pi$.

Chen \& Fried~\cite{Chen2008} used arguments similar to those of Foias et al.~\cite{Foias2001a} to predict the scaling behavior of the NS-$\alpha\beta$ model in the inertial and dissipation ranges. They showed that, like the NS-$\alpha$ model, the NS-$\alpha\beta$ model is expected to recover Kolmogorov's $k^{-5/3}$ inertial range scaling and a faster $k^{-3}$ falloff at higher wavenumbers. They offered a scaling analysis by which the wave number at which viscous dissipation becomes dominant is modified by a factor $\alpha/\beta$, implying that the inertial range of the NS-$\alpha\beta$ is extended by that factor with respect to the NS-$\alpha$ model. Kim et al.~\cite{Kim2009} studied the NS-$\alpha\beta$ model numerically. They conducted numerical simulations of forced, homogeneous and isotropic turbulence in a periodic, cubic domain and showed that, provided reasonable choices of the two parameters $\alpha$ and $\beta$, the NS-$\alpha\beta$ model, at lower resolutions, can closely approximate results obtained from highly-resolved simulations based on the NS equations. Confirming the theoretical predictions of Chen \& Fried~\cite{Chen2008}, they found that the energy spectrum predicted by the NS-$\alpha\beta$ model for $\alpha > \beta$ is more accurate than the energy spectrum from NS-$\alpha$ results at the same resolution. In recent numerical studies Kim et al.~\cite{Kim2011a} investigated the utility of the NS-$\alpha\beta$ model as a platform for spectral multigrid methods for periodic, homogeneous and isotropic turbulence. Supporting the view that the NS-$\alpha\beta$ model can approximate the NS-equations at lower resolutions, it was shown that the use of the NS-$\alpha\beta$ model at coarse grid levels and the NS equations at fine grid levels can accelerate the convergence rates compared to the convergence rates of the same spectral multigrid method based on the NS-equations only. Kim et al.~\cite{Kim2011} investigated the usability of the NS-$\alpha\beta$ model for inhomogeneous flow with non-periodic boundary conditions. They presented a similarity theory for the NS-$\alpha\beta$ model to predict the microscale, which is found to be smaller than the microscale of the NS-$\alpha$ model, leading to the assumption that the NS-$\alpha\beta$ model is able to capture smaller flow structures than the NS-$\alpha$ model, provided that $\alpha>\beta$. In numerical simulations with a finite-element method, they considered the turbulent flow in two- and three-dimensional channels past forward-backward steps. The results indicate that the physical features (in particular, eddy formation in the recirculation zone and eddy detachment) of this well-studied flow are qualitatively better captured by the NS-$\alpha\beta$ model for the choice of $\beta = \alpha /2$ than the NS-$\alpha$ model at the same resolution and same $\alpha$. These theoretical and numerical studies of the NS-$\alpha\beta$ model suggest rough guidelines for choosing the length scales $\alpha$ and $\beta<\alpha$ both usually on the order of the mesh width.

\section{The Navier--Stokes-$\alpha\beta$ model without SGS stress}

To investigate the influence of the modified viscous term in the NS-$\alpha\beta$ model isolated from the influence of the SGS stress, we formally drop the SGS stress tensor $\bftau_\alpha$ from~\eref{eq:NSalphaC2}$_1$ and arrive at the system
\begin{equation}\label{eq:NSbeta}
\left.
\begin{array}{c}
\rho \dot{\bfu} = -\Grad \overline P+\mu \triangle \bfu   
+ \mu (\alpha^2 - \beta^2) {\cal H}_\alpha^{-1} \triangle \triangle \bfu, 
\cr\noalign{\vskip4pt}
\Div \bfu = 0. 
\end{array}
\!\!
\right\}
\end{equation}
We refer to~\eref{eq:NSbeta} as the NS-$\alpha\beta*$ model. Notice that the model~\eref{eq:NSbeta} as defined here through the condition $\bftau_\alpha =\zed$ does not coincide with the limit case of $\alpha=0$ of the NS-$\alpha\beta$ model, since $\alpha$ and the inverse modified Helmholtz operator ${\cal H}_\alpha^{-1}$ still appear in the viscous term of equation~\eref{eq:NSbeta}$_1$. For a numerical study of the limit case $\alpha=0$, see Kim et al.~\cite{Kim2012}. The NS-$\alpha\beta*$ model \eref{eq:NSbeta} involves only one velocity field and its governing equation can also be seen as the NS equation supplemented with an additional Helmholtz filtered hyperviscous term. In Fourier space, the viscous term on the right-hand side of~\eref{eq:NSbeta}$_1$ is
\begin{equation}\label{eq:dissFourier}
-\mu k^2 \left( 1- \frac{\alpha^2-\beta^2}{1+\alpha^2k^2}k^2 \right) \hat{\bfu}_\bfk,
\end{equation}
where $k$ is the wavenumber and $\hat{\bfu}_\bfk$ is the corresponding complex Fourier coefficient of the velocity field $\bfu$. Notice that for $\beta=\alpha$ \eref{eq:dissFourier} reduces to the Fourier space representation $-\mu k^2\hat{\bfu}_\bfk$ of the viscous term of the NS equation. For $\beta<\alpha$ the viscous damping \eref{eq:dissFourier} monotonically decreases with increasing wavenumber $k$. Since the viscous term of~\eref{eq:NSbeta}$_1$ is identical to the viscous term of the NS-$\alpha\beta$ model, the NS-$\alpha\beta$ model is expected to have less damping at smaller scales corresponding to higher wavenumbers than the NS-$\alpha$ model. This conforms with the view that the NS-$\alpha\beta$ model compensates a possible overdamping of small-scale features in the NS-$\alpha$ model through a modified viscous term (Kim et al.~\cite{Kim2009}).

To derive the energy balance for the NS-$\alpha\beta*$ model, we first emulate the argument leading from \eref{eq:NSalphabeta}$_1$ to \eref{eq:NSabRot} to obtain an alternative version
\begin{equation}\label{eq:NSbRot}
\frac{\partial \bfu }{\partial t} - \bfu \times \C \bfu
=-\Grad \frac{\overline{P}}{\rho} + \nu [ \triangle \bfu +  (\alpha^2 - \beta^2) {\cal H}_\alpha^{-1} (\triangle \triangle \bfu)]
\end{equation}
flow equation~\eref{eq:NSbeta}$_1$. Proceeding as in the derivation of \eref{eq:NSabebalance} from \eref{eq:NSabRot}, then yields
\begin{equation}\label{eq:NSbebalance}
\dt{E_*}=\epsilon_*,
\end{equation}
with
\begin{equation}\label{eq:NSbkenergy}
E_*=\int\limits_R \half  |\bfu|^2 \dv
\end{equation}
and
\begin{equation}\label{eq:NSbkenergy2}
\epsilon_*=\int\limits_R \nu[|\bfomega|^2+
(\alpha^2 - \beta^2) \bfu \cdot {\cal H}_\alpha^{-1} (\triangle \triangle \bfu) ] \dv,
\end{equation}
where $\bfomega = \Curl \bfu$ is the filtered vorticity.

\section{Statistics of the various models}

\begin{figure}[!t]
\begin{center}
\begin{picture}(500,200)
 \put(100,0){\epsfig{file=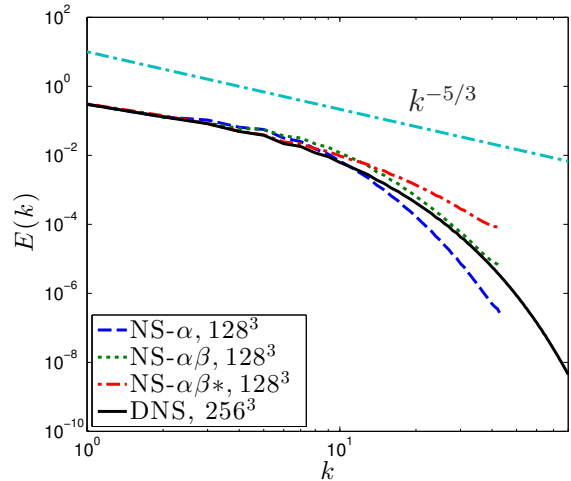,width=7.5cm}}
 \put(250,140){$k^{-5/3}$}
 \end{picture}
 \end{center}
 \caption{Energy spectra $E(k)$ of NS-$\alpha$ ($\alpha=1/8$) and NS-$\alpha\beta$ ($\alpha=1/8, \beta=1/12$) at $128^3$ resolution, along with those of NS-$\alpha\beta*$ model ($\alpha=1/8, \beta=1/12$) at $128^3$ resolution and DNS at $256^3$ resolution.}
 \label{espE}
\end{figure}

We next discuss numerical results of the NS-$\alpha\beta$ model, the NS-$\alpha$ model, and the NS-$\alpha\beta*$ model in comparison to results from high resolution DNS. Direct numerical simulations using~\eref{eq:NSalphabeta}$_{2,3}$ and~\eref{eq:NSabRot} for the NS-$\alpha$ and NS-$\alpha\beta$ models and using~\eref{eq:NSbeta}$_{3}$ and~\eref{eq:NSbRot} for the NS-$\alpha\beta*$ model are performed for three-dimensional homogeneous and isotropic turbulent flows with periodic boundary conditions.
We use the pseudospectral method for the spatial discretization and a second-order Adams--Bashforth scheme for time advancement. We employ a forcing scheme developed by Chen et al.~\cite{Chen1993} that maintains constant energy in the first two wavenumber shells and, thus, extends the inertial range. The ratio between the energy contained in the first two wavenumber shells is chosen to be compatible with the $k^{-5/3}$ inertial range scaling. Unless otherwise indicated, all the simulations are performed using a (dimensionless) viscosity of $\nu=0.005$ corresponding to the Reynolds number $Re=200$.

All results are averaged over several large-eddy turnover times. We start collecting data once the flow reaches a statistically stationary state, which occurs after approximately ten large-eddy turnover times. In the statistically stationary regime, we collect at least ten data sets over approximately thirteen large-eddy turnover times and then average those data sets. 
 
Numerical studies of regularization models available in the literature exhibit important differences on the strategy of comparison of results and evaluation of the quality of a certain model to produce approximations to DNS results. We next briefly discuss the differences and describe our strategy. Most numerical studies of large-eddy simulation (LES) models aim at comparing the predictions of a given LES model at a significantly lower resolution with a reference solution obtained from experiments or DNS data at a higher resolution. In many such studies, results are not compared directly to DNS data at higher resolution, but instead a reference solution is usually constructed by filtering high resolution DNS data and projecting the filtered data on the low resolution LES grid, as explained by Meneveau \& Katz~\cite{Meneveau2000}. For the evaluation and comparison of regularization models there has been examples of using unfiltered DNS data at higher resolution as reference solution. Chen et al.~\cite{Chen1999b}, Mohseni et al.~\cite{Mohseni2003}, and Pietarila Graham et al.~\cite{PietarilaGraham2007, PietarilaGraham2008} used this strategy in their numerical studies of the NS-$\alpha$ model. On the other side, Geurts et al.~\cite{Geurts2008} adapted the LES strategy of using filtered and projected DNS data in their numerical studies of different regularization models. The general LES strategy of resolving only the large, energy containing scales while modeling the smaller scales including the dissipation range is based on the assumption that a separation of scales, normally only found  in high or very high Reynolds number flows prevails. However, work by Pietarila Graham et al.\cite{PietarilaGraham2007, PietarilaGraham2008} has demonstrated, that there are several difficulties with different regularization models at higher Reynolds numbers. Also, as opposed to LES, the NS-$\alpha \beta$ model aims to model a larger range of the energy spectrum, including part of the small scale motion in the dissipation range and, thus, comparing to filtered and projected DNS data at higher resolution is not suited to our objectives. 
We therefore adopt the strategy of comparing results of the NS-$\alpha$ model, the NS-$\alpha\beta$ model, and the NS-$\alpha\beta*$ model at lower resolution to unfiltered DNS results at higher spatial resolution.

\begin{figure*}[!t]
\begin{center}
\begin{picture}(450,200)
 \put(0,0){\epsfig{file=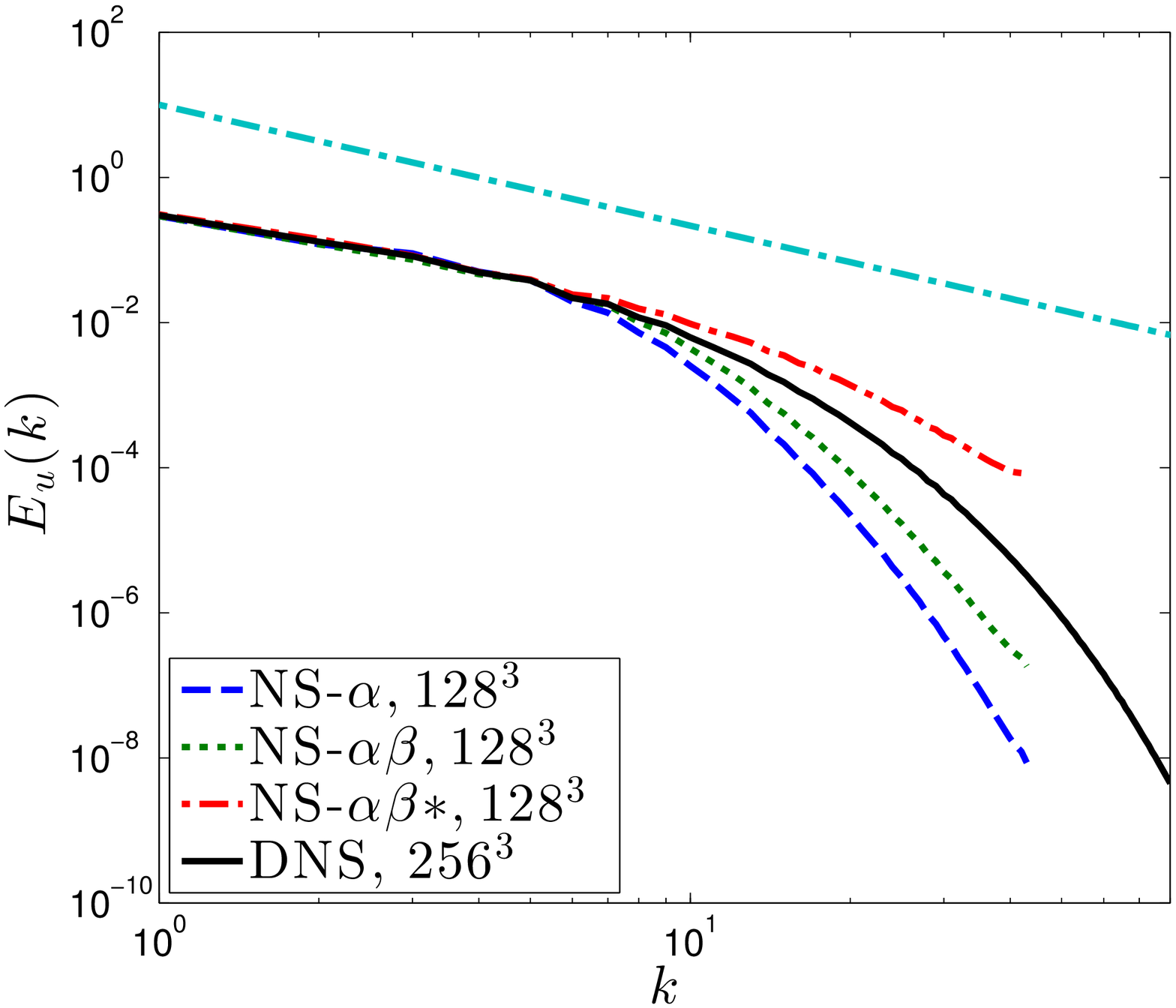,width=7.5cm}}
 \put(220,0){\epsfig{file=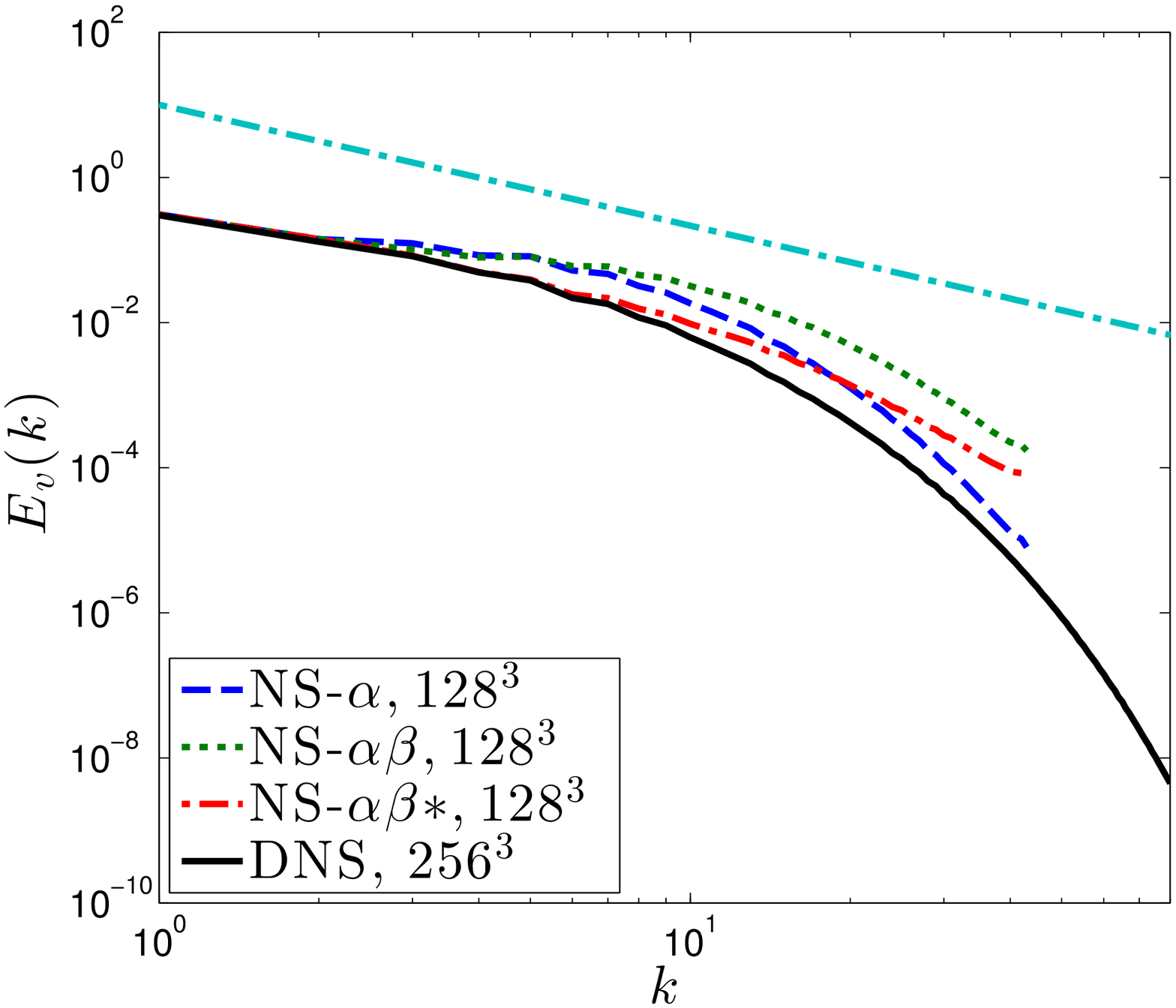,width=7.5cm}}
  \put(150,140){$k^{-5/3}$}
  \put(370,140){$k^{-5/3}$}
 \put(115,-15){(a)}
 \put(330,-15){(b)}
 \end{picture}
 \end{center}
 \caption{Alternative energy spectra (a) $E_u(k)$ and (b) $E_v(k)$ for NS-$\alpha$ ($\alpha=1/8$) and NS-$\alpha\beta$ ($\alpha=1/8, \beta=1/12$) at $128^3$ resolution compared to energy spectra of NS-$\alpha\beta*$ ($\alpha=1/8, \beta=1/12$) at $128^3$ resolution and DNS results at $256^3$ resolution.} 
 \label{espEuv}
\end{figure*}

\vspace{24pt}
\subsection{Energy spectra}

In this section, we study energy spectra for the NS-$\alpha\beta$ model along with DNS data at higher resolution. Specifically, we highlight differences between the norms commonly used for the discussion of results from regularization models, namely
\begin{equation}\label{eq:norms}
\displaystyle
 E= \frac{1}{2}\int\limits_{R} \bfv\cdot\bfu \dv,\qquad  
 \displaystyle
 E_u =\frac{1}{2} \int\limits_{R} |\bfu|^2 \dv, \qquad E_v =\frac{1}{2} \int\limits_{R} |\bfv|^2 \dv.
\end{equation}
Introducing the wavenumber $k$, we denote by $E(k)$, $E_u(k)$, and $E_v(k)$ the respective spectra computed from the above norms. Since $\bfv=(1-\alpha^2\Delta)\bfu$, \eref{eq:norms}$_1$ is equivalent to~\eref{eq:NSabkenergy}$_1$ and, thus, the only natural norm for the NS-$\alpha$ and NS-$\alpha\beta$ models, arising in the energy balance~\eref{eq:NSabebalance}. Adopting LES terminology, the norm \eref{eq:norms}$_2$ and the corresponding spectrum $E_u(k)$, involving only the filtered velocity $\bfu$, can be interpreted as the spectrum of the resolved kinetic energy. 
In view of~\eref{eq:NSbkenergy}, the natural energy norm of the  NS-$\alpha\beta*$ model is given by~\eref{eq:norms}$_2$. 

To begin, we display the natural energy spectra of the NS-$\alpha\beta$ and NS-$\alpha$ models at $128^3$ resolution compared with energy spectra of the NS-$\alpha\beta*$ model at $128^3$ resolution and DNS results at $256^3$ resolution in Figure~\ref{espE}. Plots are provided for the NS-$\alpha$ ($\alpha=1/8$) and the NS-$\alpha\beta$ ($\alpha=1/8, \beta=1/12$) as well as the NS-$\alpha\beta*$ model ($\alpha=1/8, \beta=1/12$). 
In the inertial range, the natural energy spectra for both NS-$\alpha$ and NS-$\alpha\beta$ follow Kolmogorov's $-5/3$ law and are close to the spectrum obtained from DNS. Further, the natural energy spectrum of the NS-$\alpha\beta$ model closely approximates the energy spectrum of the DNS results in both the inertial and dissipation ranges. This result is consistent with the findings of Kim et al.~\cite{Kim2009}. It indicates that, for $\beta<\alpha$, the NS-$\alpha\beta$ model provides more energy at smaller scales than in the NS-$\alpha$ model. 
Compared to the natural energy spectra of the NS-$\alpha$ and NS-$\alpha\beta$ models, as well as the DNS energy spectrum, the  spectrum of the NS-$\alpha\beta*$ model includes excess energy at the intermediate and small scales. This result is consistent with the nature of the viscous term~\eref{eq:dissFourier} of the NS-$\alpha\beta*$ model, which decreases with increasing wavenumber, providing less damping, as discussed above. We thus confirm the view that the NS-$\alpha\beta$ model may be seen as a combination of the NS-$\alpha$ model and the NS-$\alpha\beta*$ model, a combination that contains the SGS stress tensor terms of the NS-$\alpha$ model and the modified viscous term of the NS-$\alpha\beta*$ model and which, granted a choice of parameters $\beta < \alpha$, allows for less damping at the smaller scales. Notice that the energy spectrum of the NS-$\alpha\beta*$ model does not provide good approximations to DNS energy spectra at higher resolution and, thus, should not be viewed as a suitable regularization model for fluid turbulence. Rather, it should be viewed as an additional test case that reveals the influence of the tuning parameter $\beta$ in the NS-$\alpha\beta$ model. In the following we provide and discuss statistical results of the NS-$\alpha\beta*$ model, along with the corresponding results of the NS-$\alpha$ and NS-$\alpha\beta$ models, as well as DNS results at higher resolution.

The alternative spectra of $E_u(k)$ and $E_v(k)$ of the NS-$\alpha$ and NS-$\alpha\beta$ models are shown in Figure~\ref{espEuv}. For both the NS-$\alpha$ and NS-$\alpha\beta$ models, the resolved kinetic energy spectrum $E_u(k)$  follows Kolmogorov's $-5/3$ law in the inertial range and is indistinguishable from the energy spectrum of the DNS results, as shown in panel (a) of Figure~\ref{espEuv}. 
However, the spectrum $E_u(k)$ of the NS-$\alpha$ and NS-$\alpha\beta$ models significantly deviates from the DNS energy spectrum in the dissipation range. Throughout the higher wavenumbers, the resolved kinetic energy spectrum of the NS-$\alpha\beta$ model is closer to the DNS energy spectrum than the corresponding resolved kinetic energy spectrum of the NS-$\alpha$ model. That is, even the filtered NS-$\alpha\beta$ velocity field contains more small-scale features than the filtered NS-$\alpha$ model. 
The energy spectra $E_v(k)$ obtained from the integral~\eref{eq:norms}$_3$ are shown in panel (b) of Figure~\ref{espEuv}. Neither the NS-$\alpha$ nor the NS-$\alpha\beta$ results follow the $k^{-5/3}$ law in the inertial range. Further, both models significantly overpredict the spectrum obtained from DNS. Strikingly, these inconsistencies are pervasive in the sense that they occur in both the inertial and the dissipation ranges. Although the  NS-$\alpha$ results are closer to the DNS results than are the NS-$\alpha\beta$ results, neither of these models provides a good approximation to the DNS results.

These results indicate that the natural energy spectrum $E(k)$ for the NS-$\alpha$ and NS-$\alpha\beta$ model shows good agreement with that of the DNS results. Importantly, the resolved kinetic energy spectrum of the filtered velocity fields provides a good approximation to the DNS results, with the NS-$\alpha\beta$ model showing better agreement than the NS-$\alpha$ model. In contrast, the spectrum of the unfiltered velocity field does not provide a suitable approximation to the DNS results because of the significant overprediction of energy contained at intermediate and high wavenumbers. Consequently, whereas the filtered velocity field has physically more reasonable statistics in spectral space, the unfiltered velocity field exhibits unphysical statistics in spectral space. 
In the following, we confirm this observation by comparing the probability density functions (PDFs), structure functions, and flatness factors of the NS-$\alpha\beta$ and NS-$\alpha$ models with those of the NS-$\alpha\beta*$ model and higher resolution DNS results.

\begin{figure*}[!t]
\begin{center}
\begin{picture}(450,470)
 \put(20,310){\epsfig{file=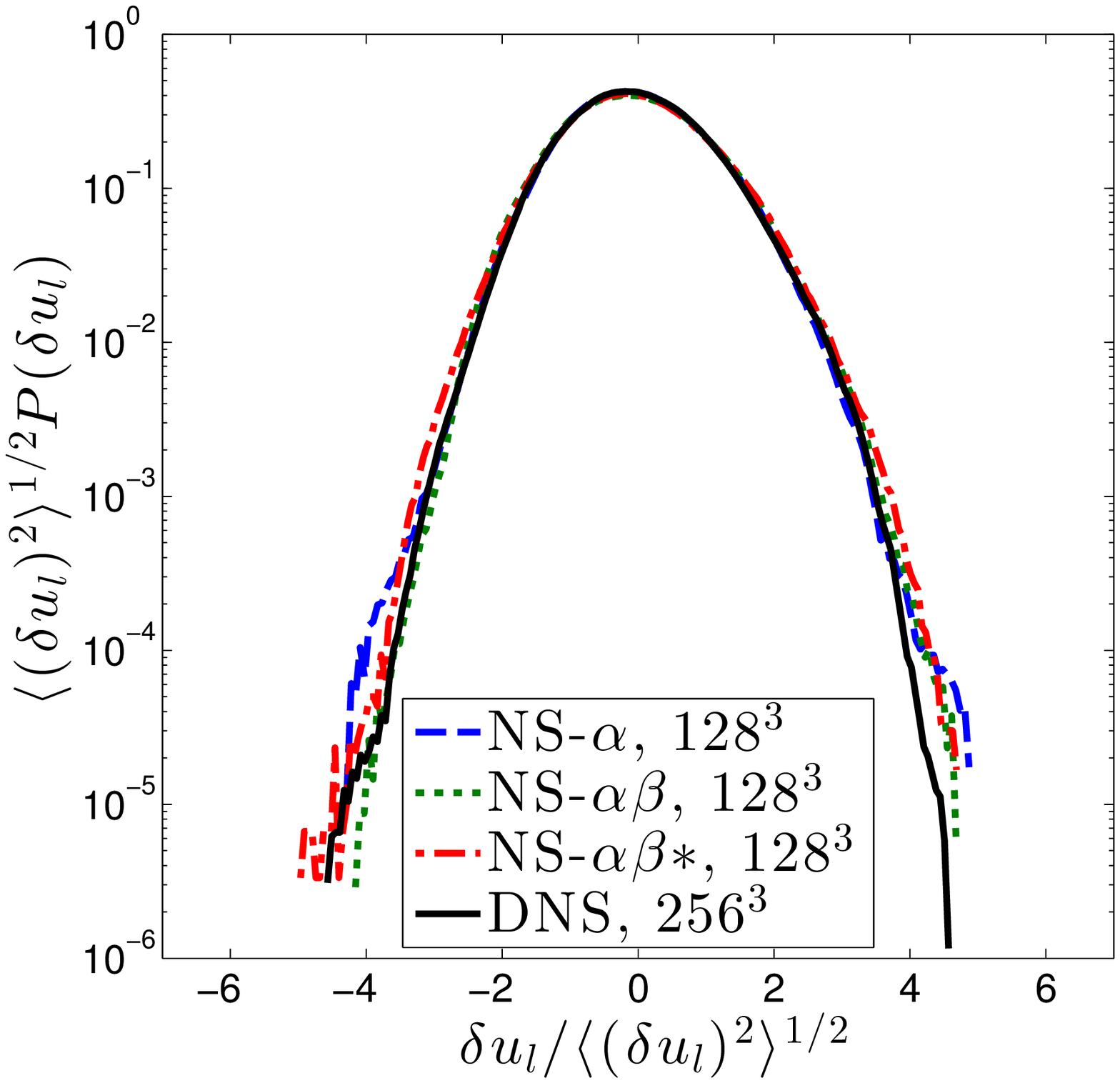,width=6.0cm, height=5.5cm}}
 \put(20,150){\epsfig{file=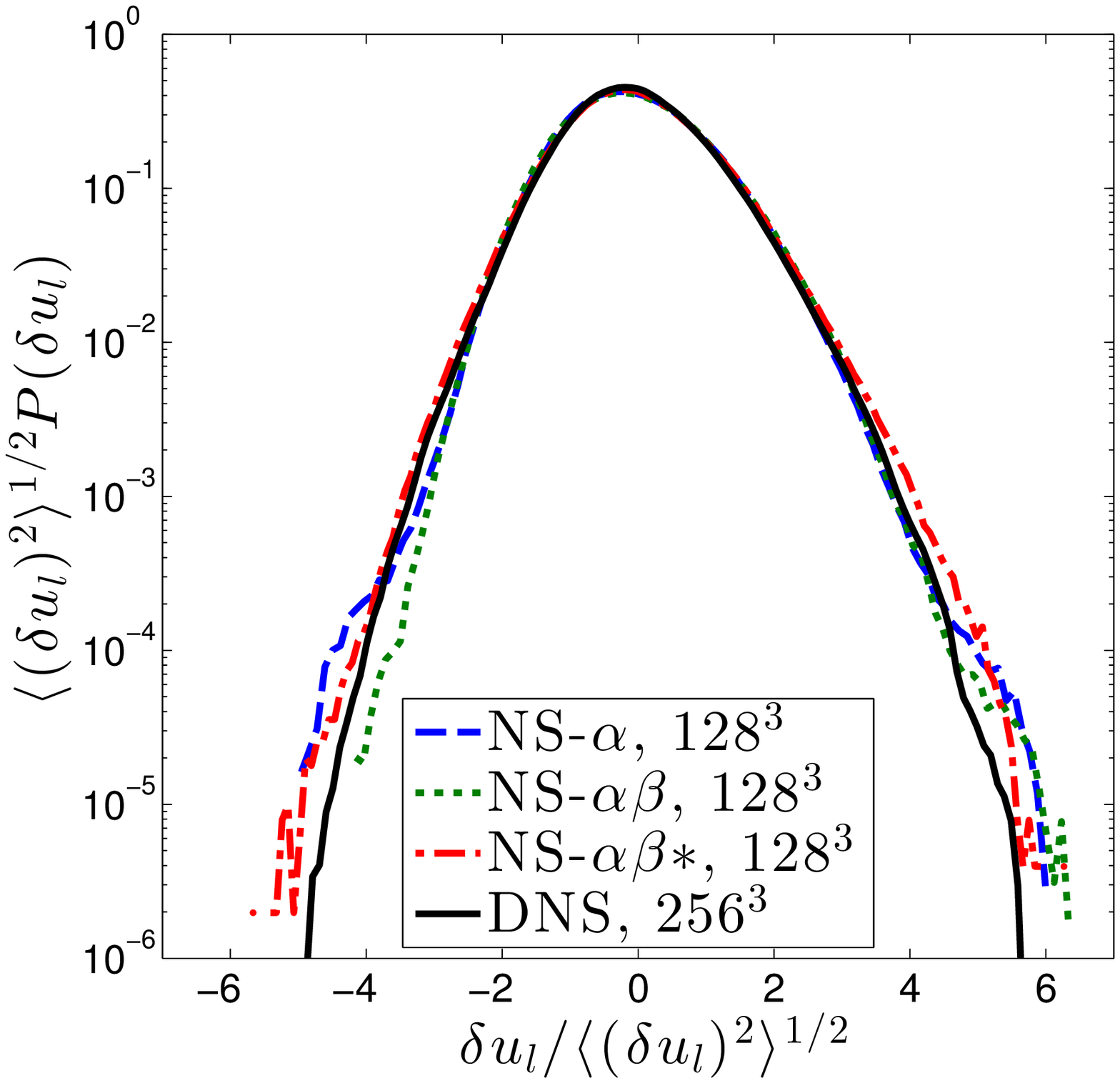,width=6.0cm, height=5.5cm}}
 \put(20,-10){\epsfig{file=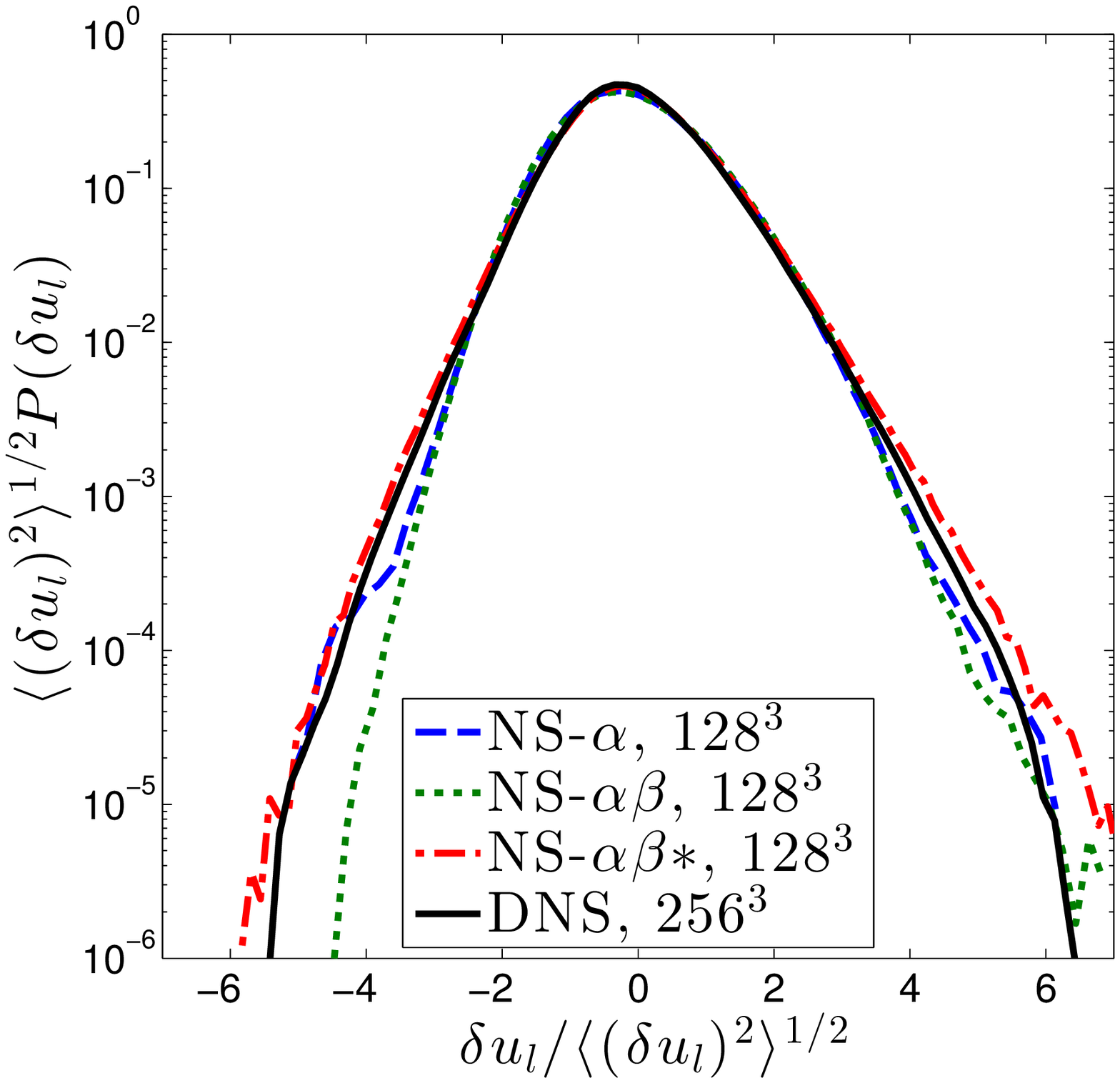,width=6.0cm, height=5.5cm}}
 
 \put(230,310){\epsfig{file=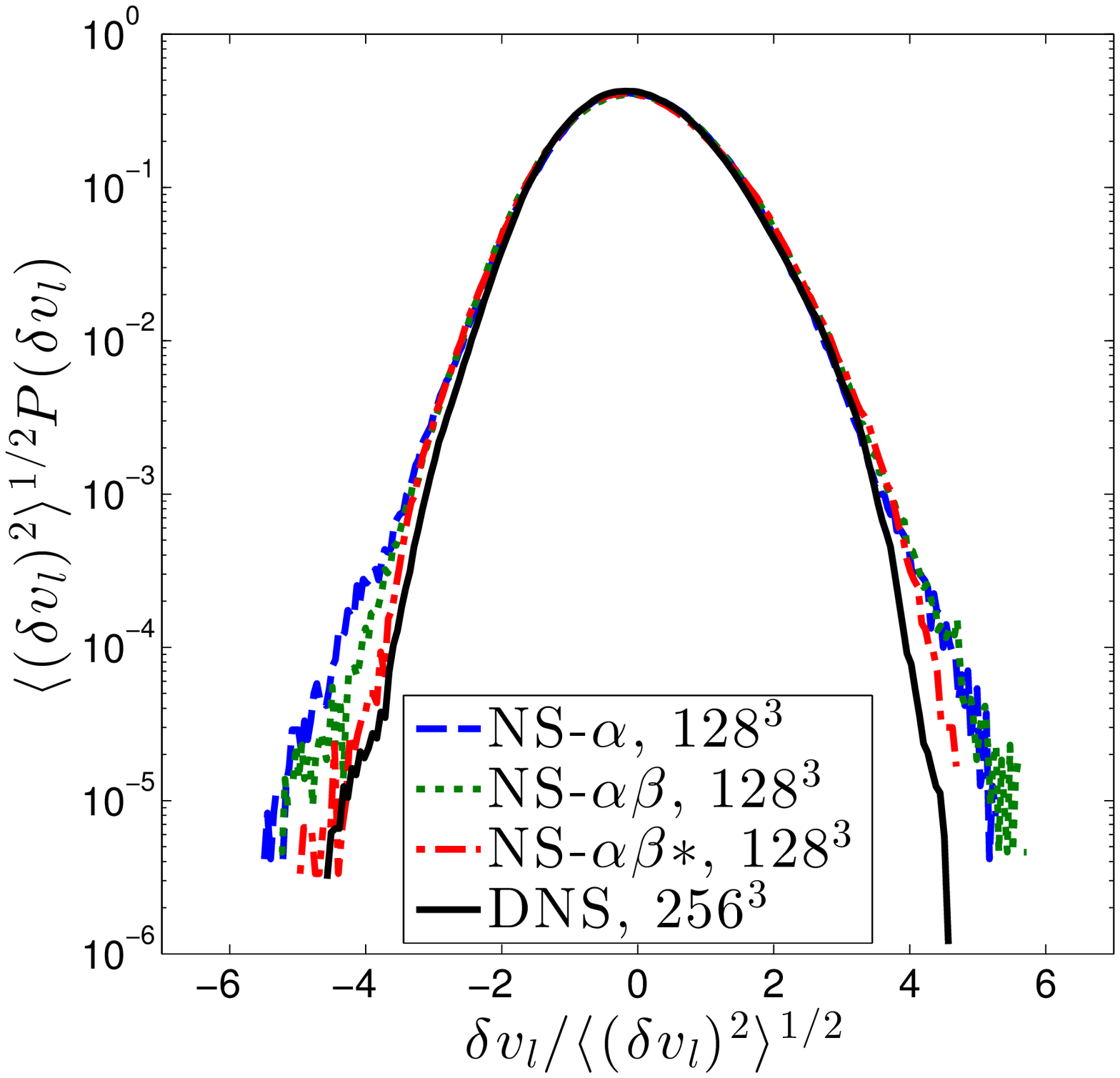,width=6.0cm, height=5.5cm}}
 \put(230,150){\epsfig{file=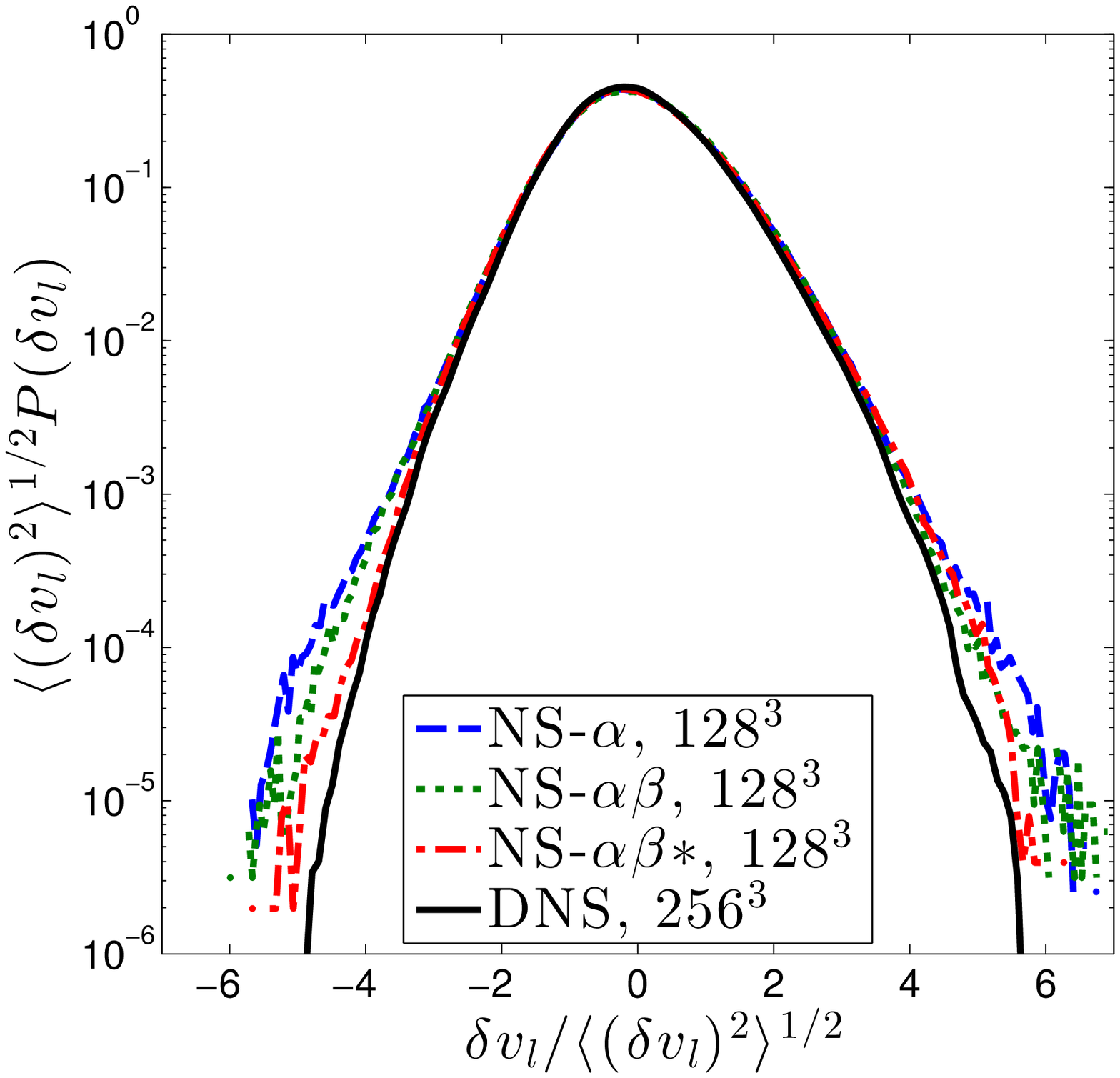,width=6.0cm, height=5.5cm}} 
 \put(230,-10){\epsfig{file=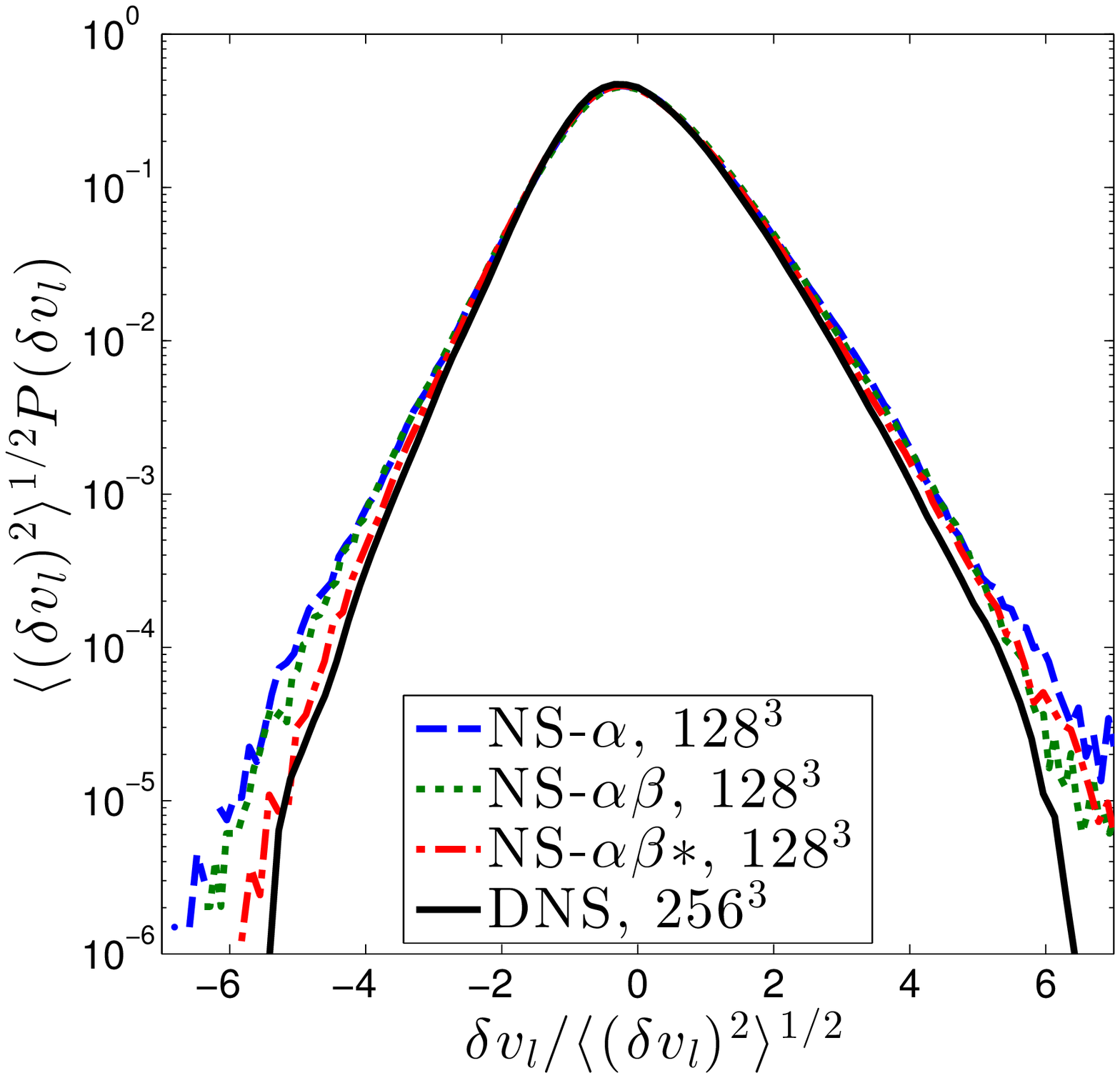,width=6.0cm, height=5.5cm}}

\put(50,450){\footnotesize(a) $r = 5/32$}
\put(50,290){\footnotesize(b) $r = 1/16$}
\put(50,130){\footnotesize(c) $r = 1/32$}

\put(260,450){\footnotesize(d) $r = 5/32$}
\put(260,290){\footnotesize(e) $r = 1/16$}
\put(260,130){\footnotesize(f) $r = 1/32$}

\end{picture}
\end{center}
 \caption{Normalized PDFs $P(\delta u_l)$ and $P(\delta v_l)$ of the filtered velocity increment $\delta u_l$ and the unfiltered velocity increment $\delta v_l$ for NS-$\alpha$ ($\alpha=1/8$) and NS-$\alpha\beta$ ($\alpha=1/8$, $\beta=1/12$) compared to NS-$\alpha\beta*$ ($\alpha=1/8, \beta=1/12$) at $128^3$ resolution and DNS at $256^3$ resolution with different normalized separation distances $r= 5/32$, $r= 1/16$, and $r= 1/32$.}
\label{PDF1n}
\end{figure*}
\begin{figure*}[!t]
\begin{center}
\begin{picture}(450,470)
 \put(20,310){\epsfig{file=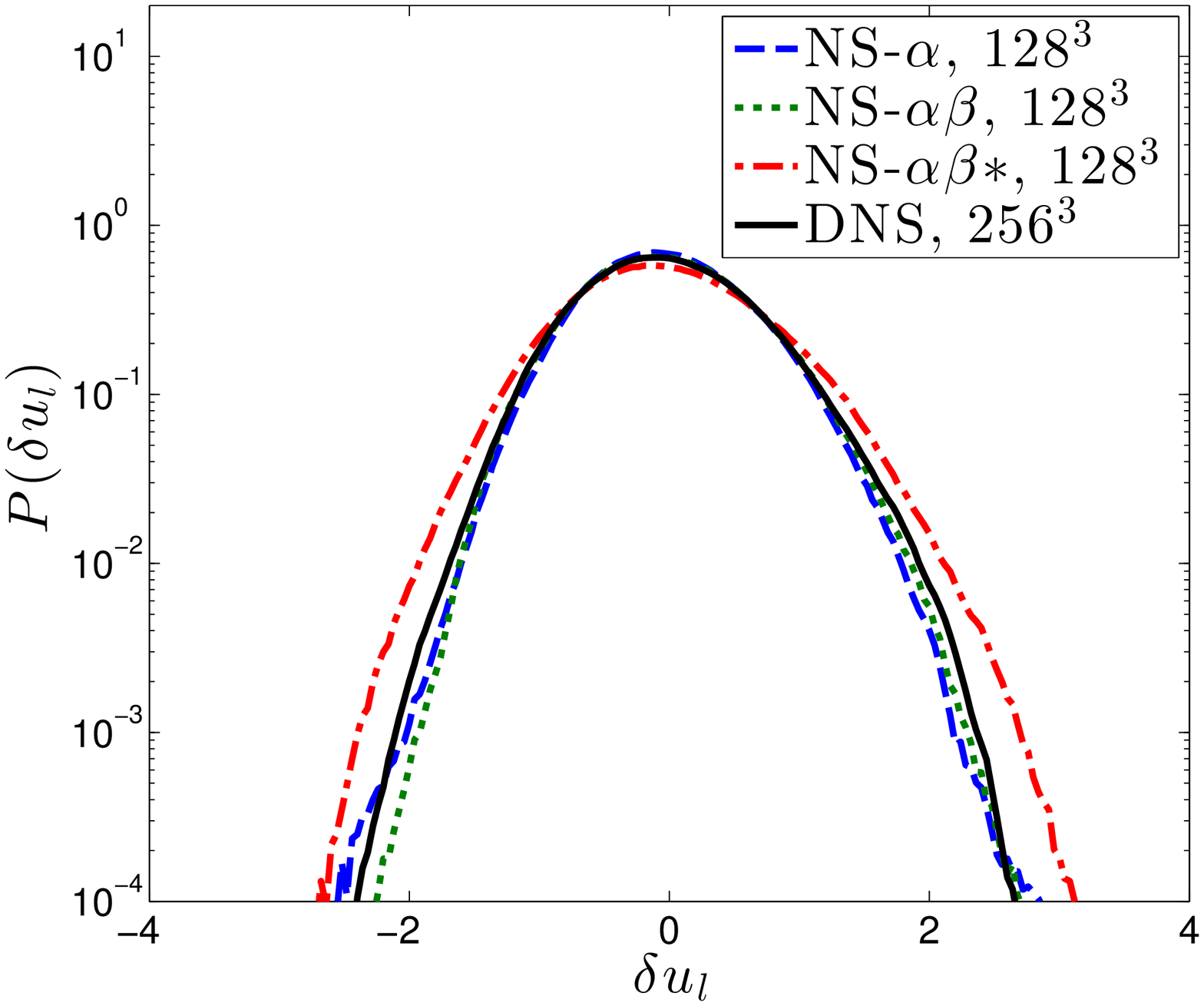,width=6.5cm, height=5.5cm}}
 \put(20,150){\epsfig{file=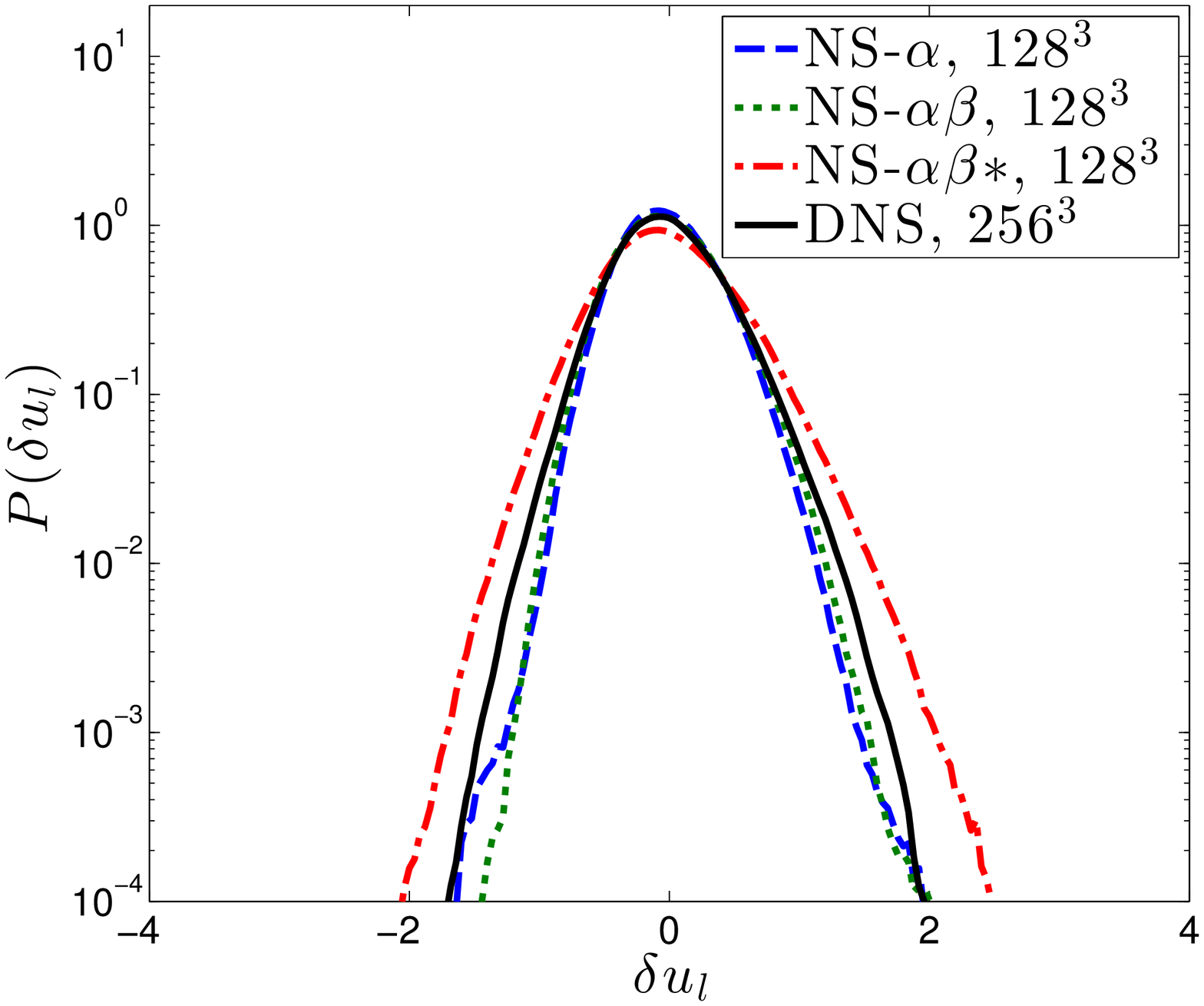,width=6.5cm, height=5.5cm}}
 \put(20,-10){\epsfig{file=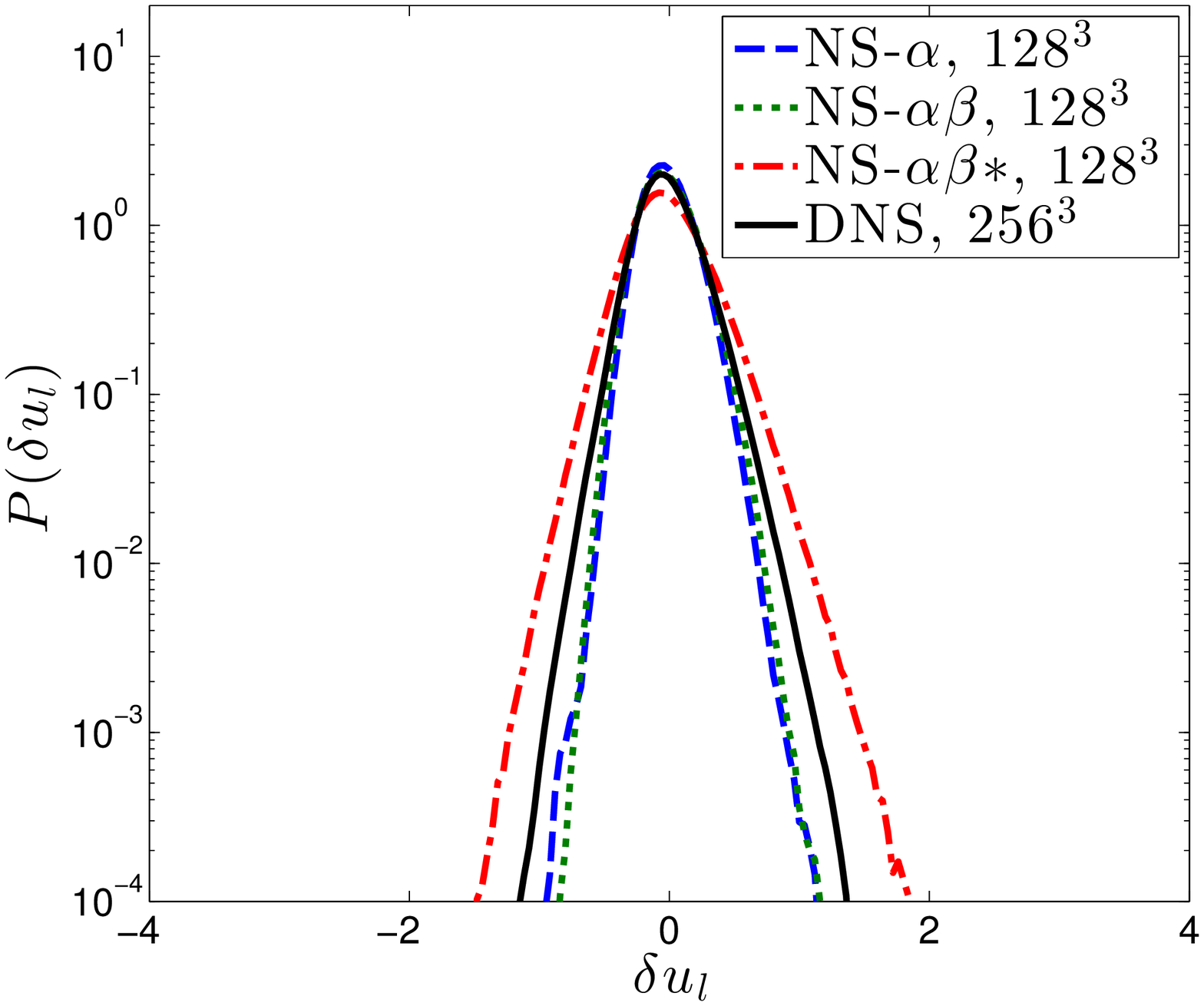,width=6.5cm, height=5.5cm}}
 
 \put(230,310){\epsfig{file=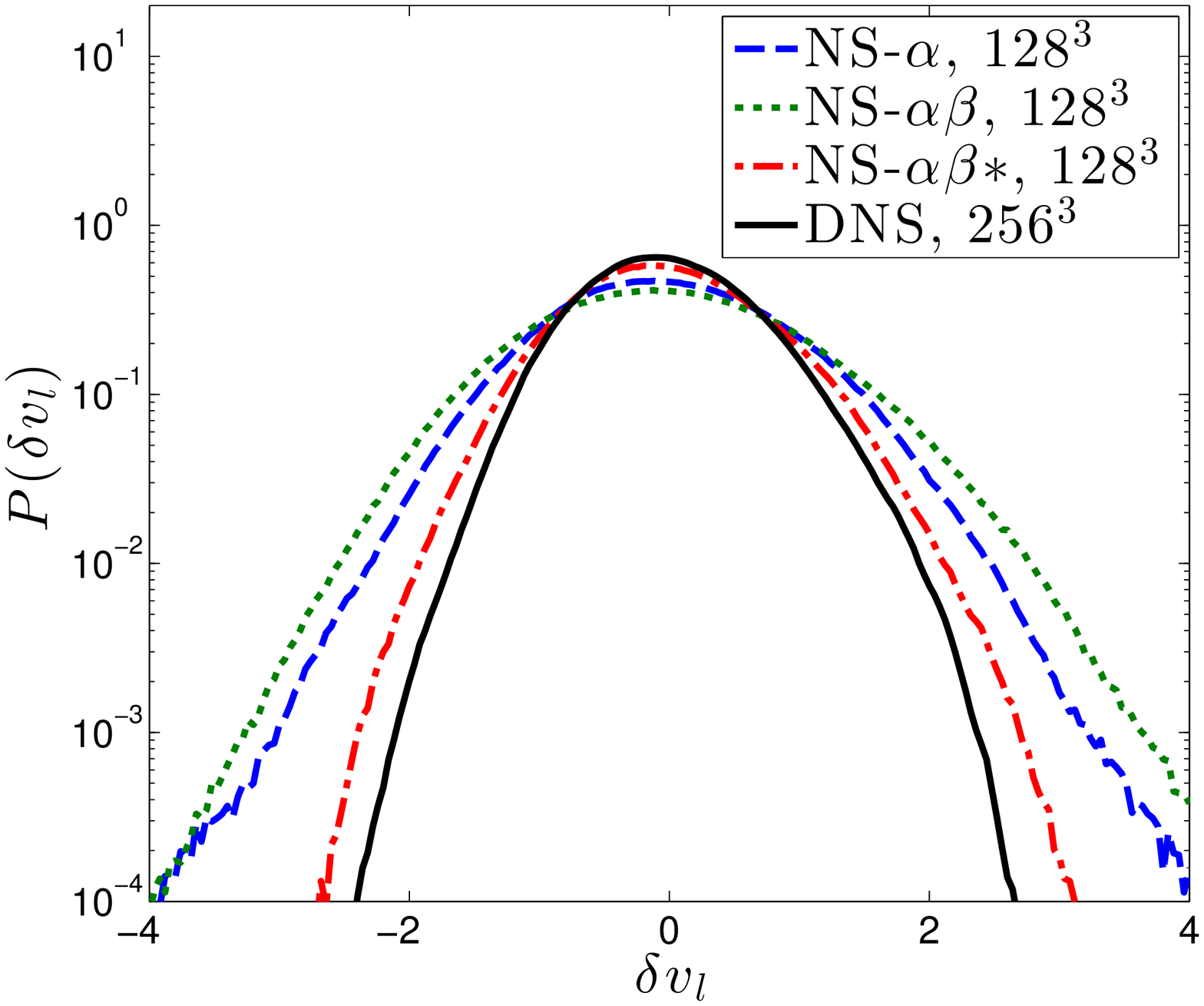,width=6.5cm, height=5.5cm}}
 \put(230,150){\epsfig{file=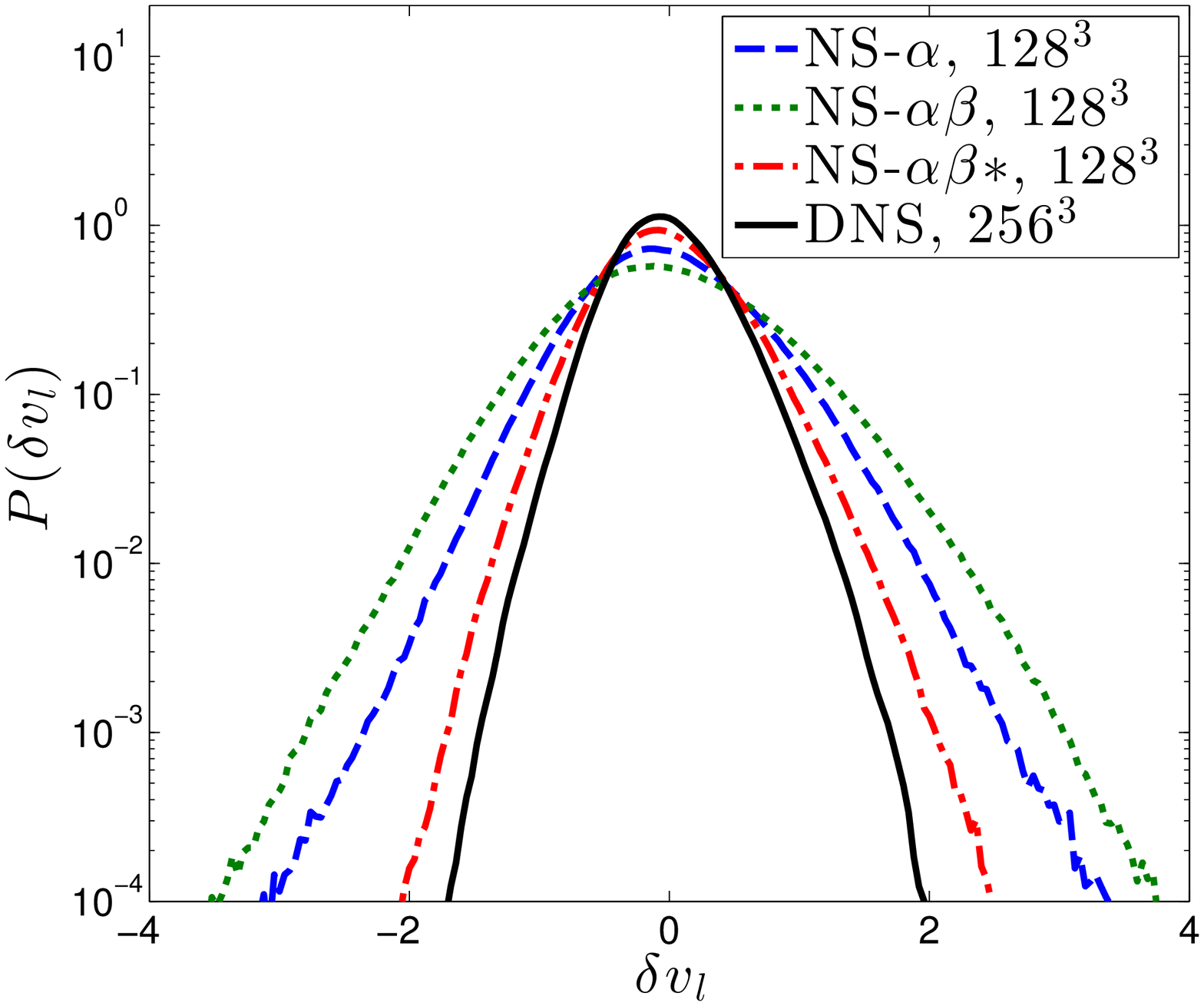,width=6.5cm, height=5.5cm}} 
 \put(230,-10){\epsfig{file=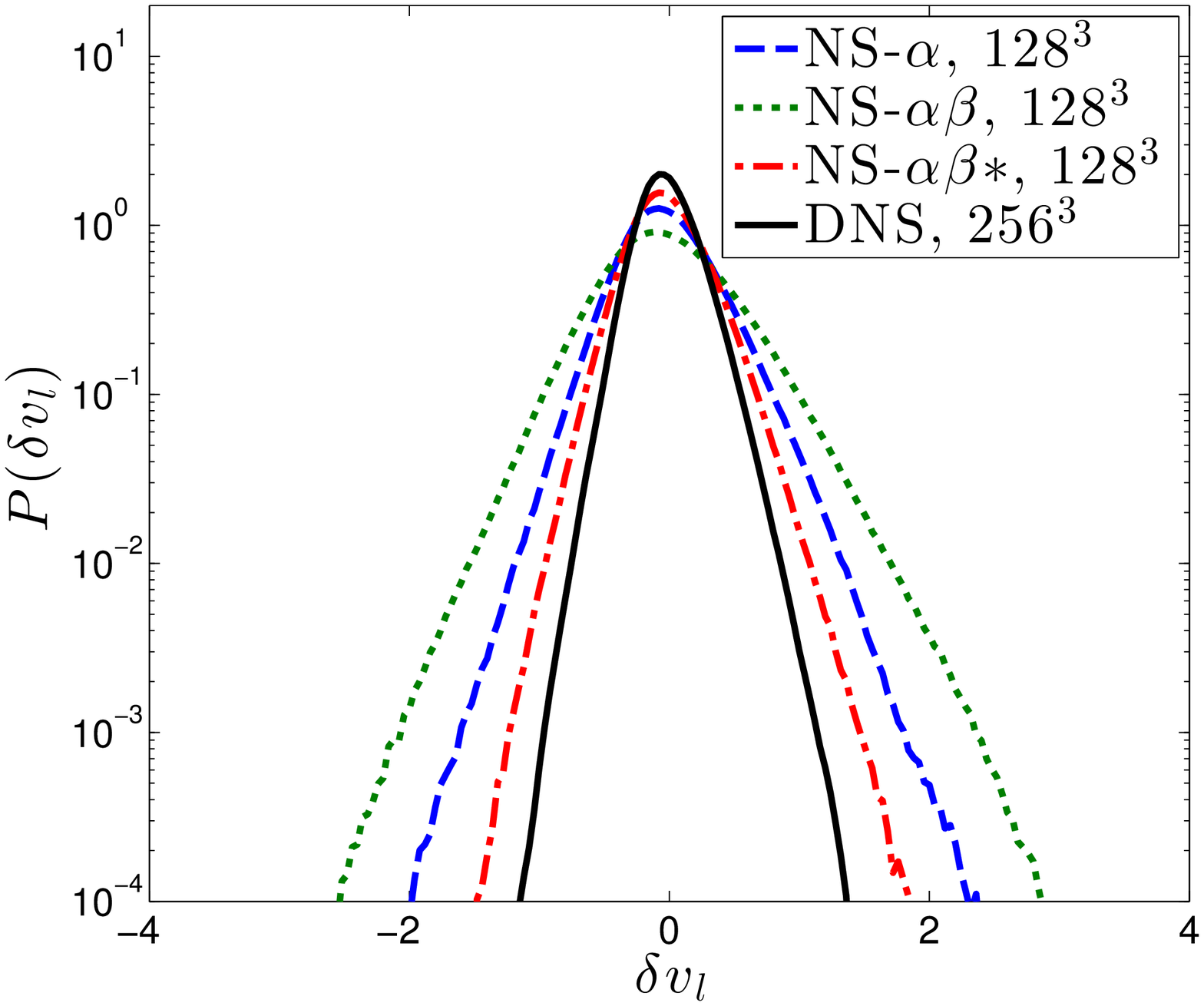,width=6.5cm, height=5.5cm}}

\put(50,450){\footnotesize(a) $r = 5/32$}
\put(50,290){\footnotesize(b) $r = 1/16$}
\put(50,130){\footnotesize(c) $r = 1/32$}

\put(260,450){\footnotesize(d) $r = 5/32$}
\put(260,290){\footnotesize(e) $r = 1/16$}
\put(260,130){\footnotesize(f) $r = 1/32$}

\end{picture}
\end{center}
 \caption{PDFs of the filtered velocity increment $\delta u_l$ and the unfiltered velocity increment $\delta v_l$ for NS-$\alpha$ ($\alpha=1/8$) and NS-$\alpha\beta$ ($\alpha=1/8$, $\beta=1/12$) compared to NS-$\alpha\beta*$ ($\alpha=1/8, \beta=1/12$) at $128^3$ resolution and DNS at $256^3$ resolution with different normalized separation distances $r= 5/32$, $r= 1/16$, and $r= 1/32$.}
\label{PDF1}
\end{figure*}

\begin{figure*}[!t]
\begin{center}
\begin{picture}(450,470)
 \put(20,310){\epsfig{file=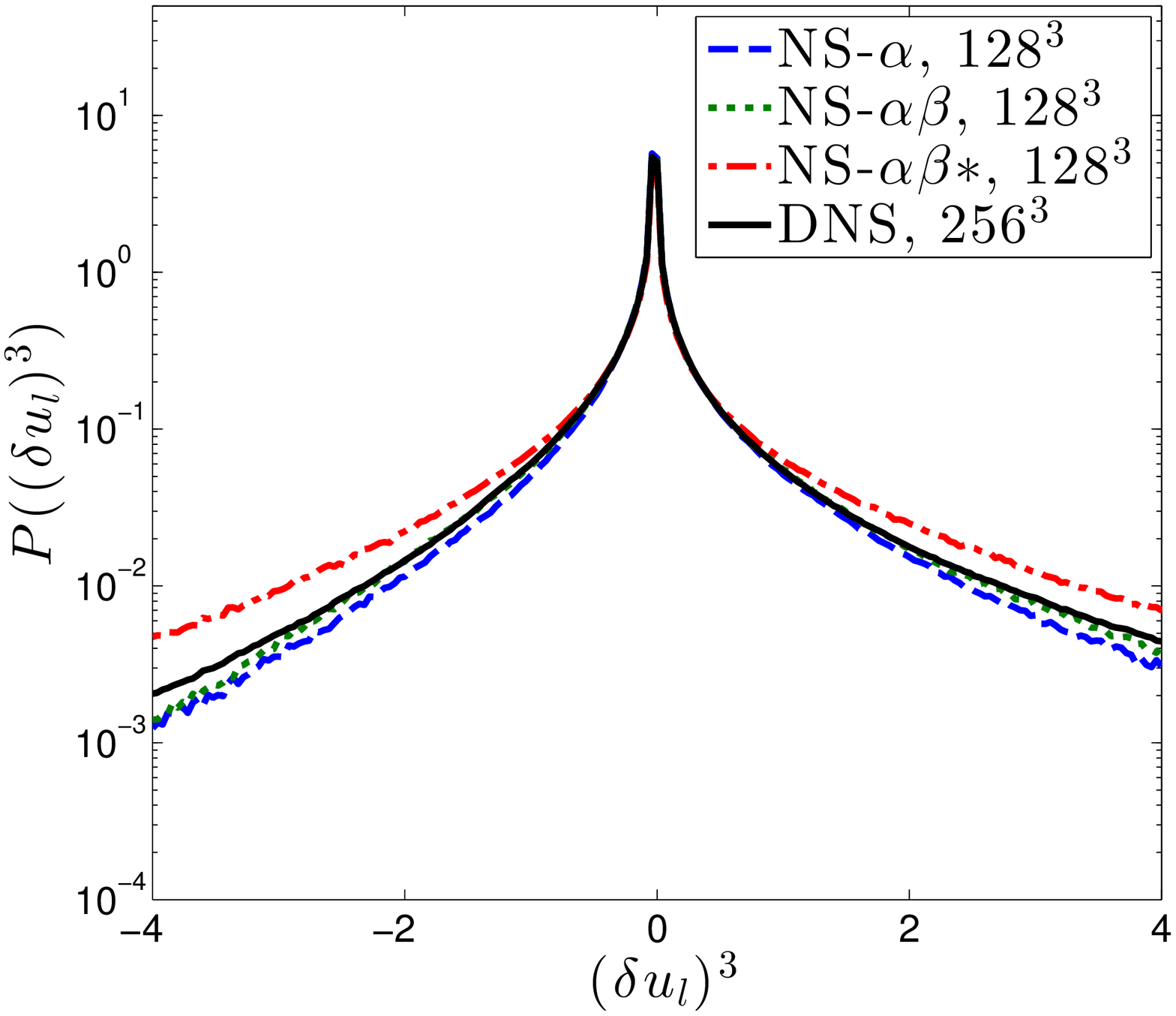,width=6.5cm, height=5.5cm}}
 \put(20,150){\epsfig{file=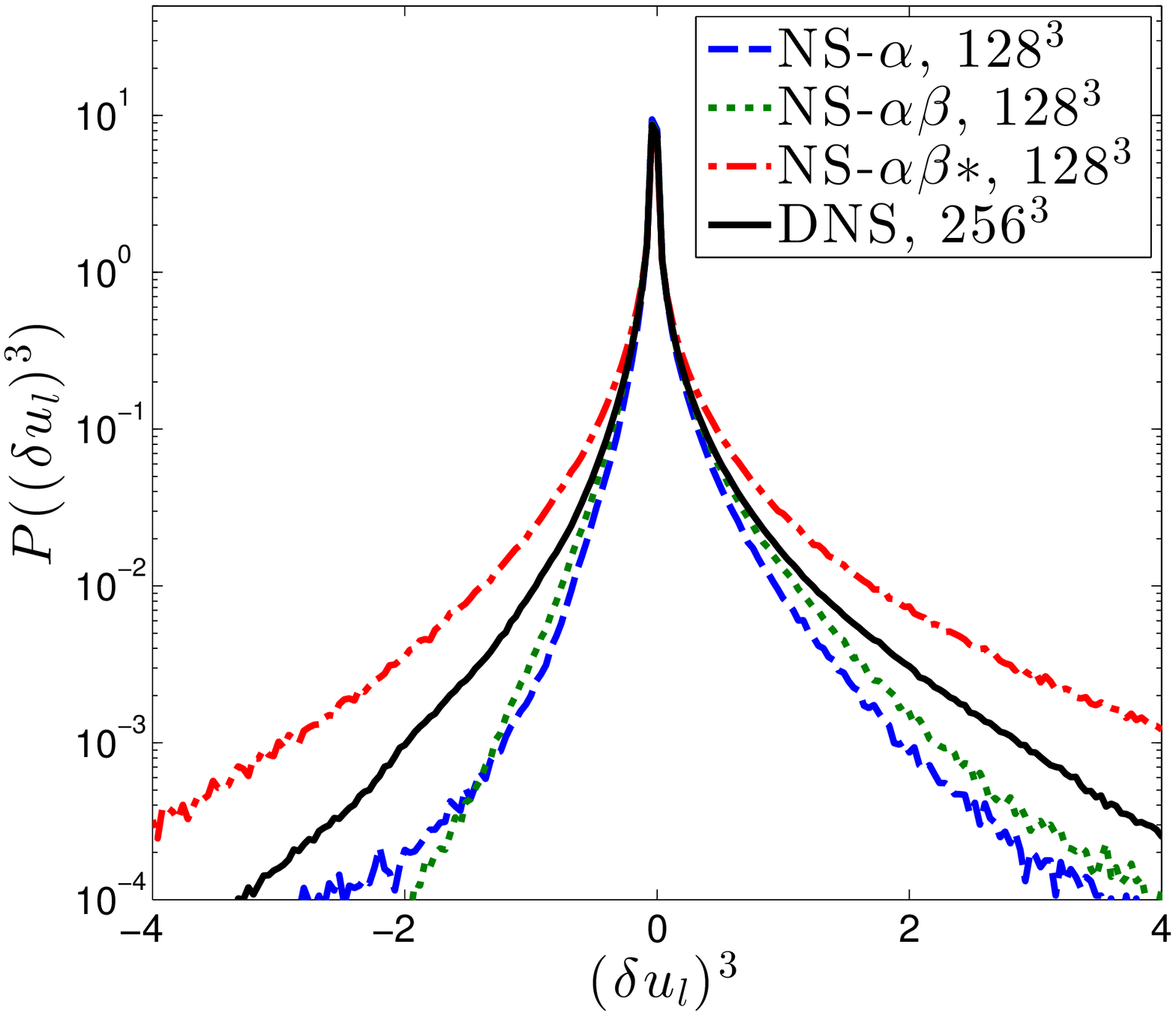,width=6.5cm, height=5.5cm}}
 \put(20,-10){\epsfig{file=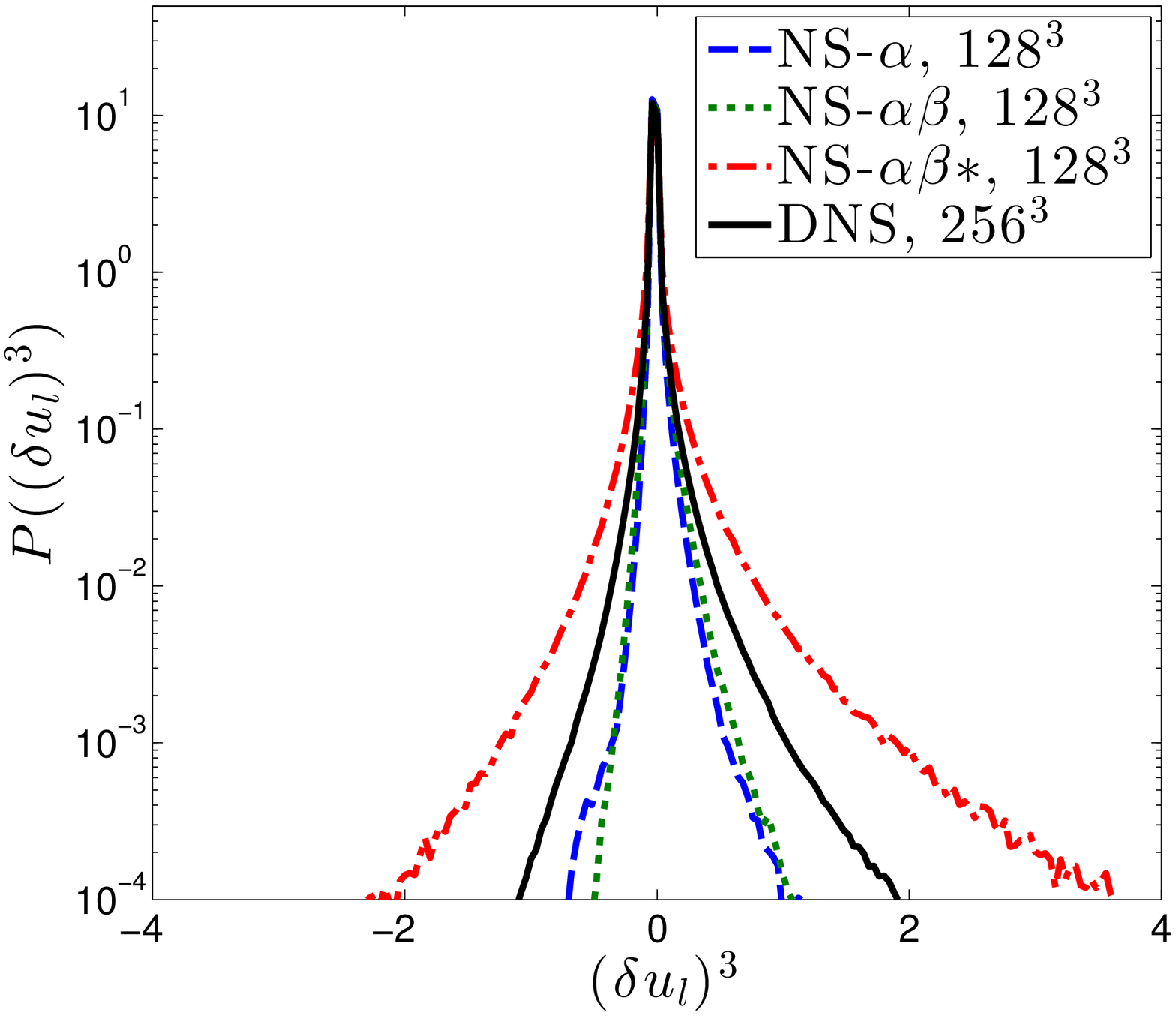,width=6.5cm, height=5.5cm}}
 
 \put(230,310){\epsfig{file=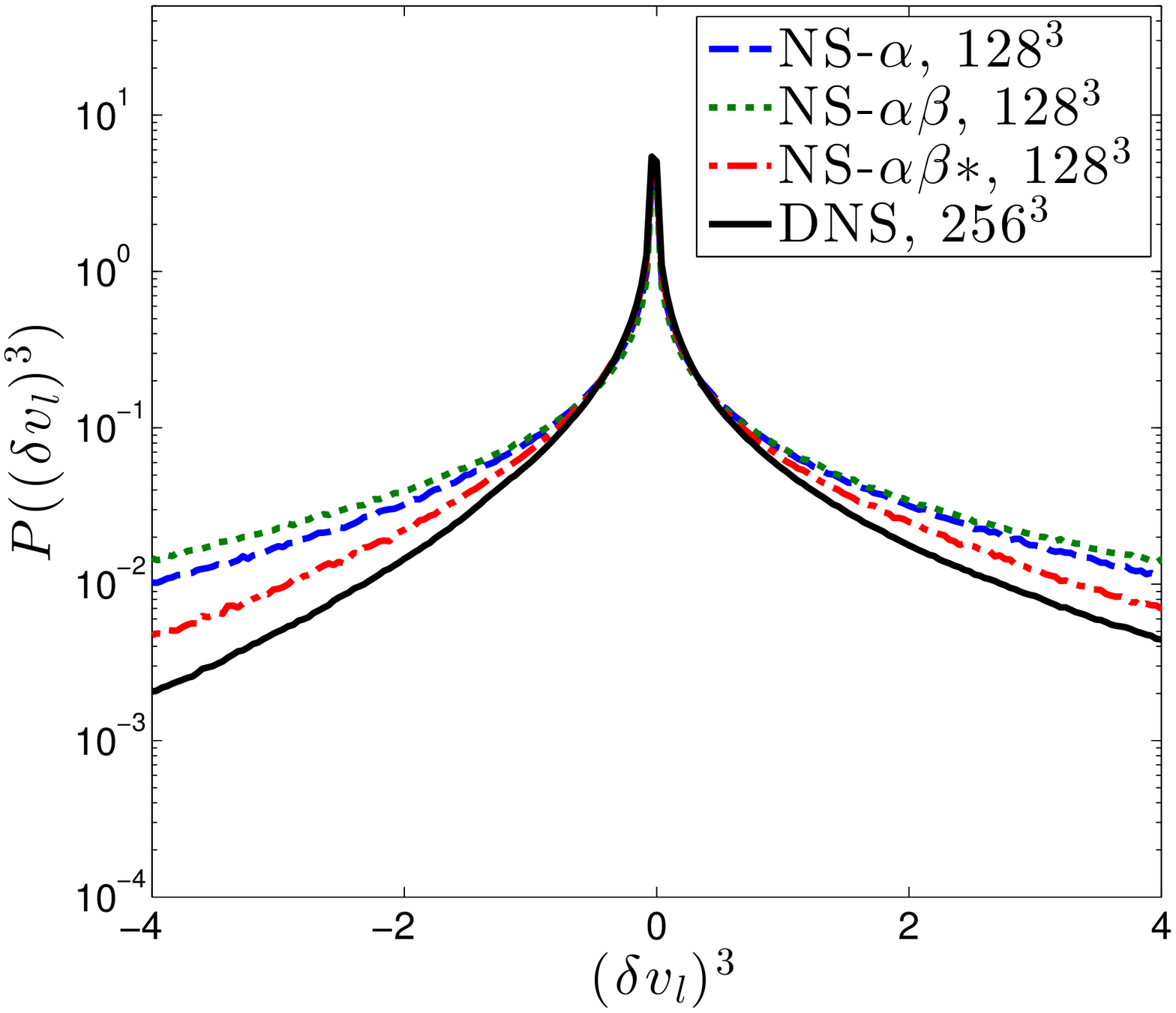,width=6.5cm, height=5.5cm}}
 \put(230,150){\epsfig{file=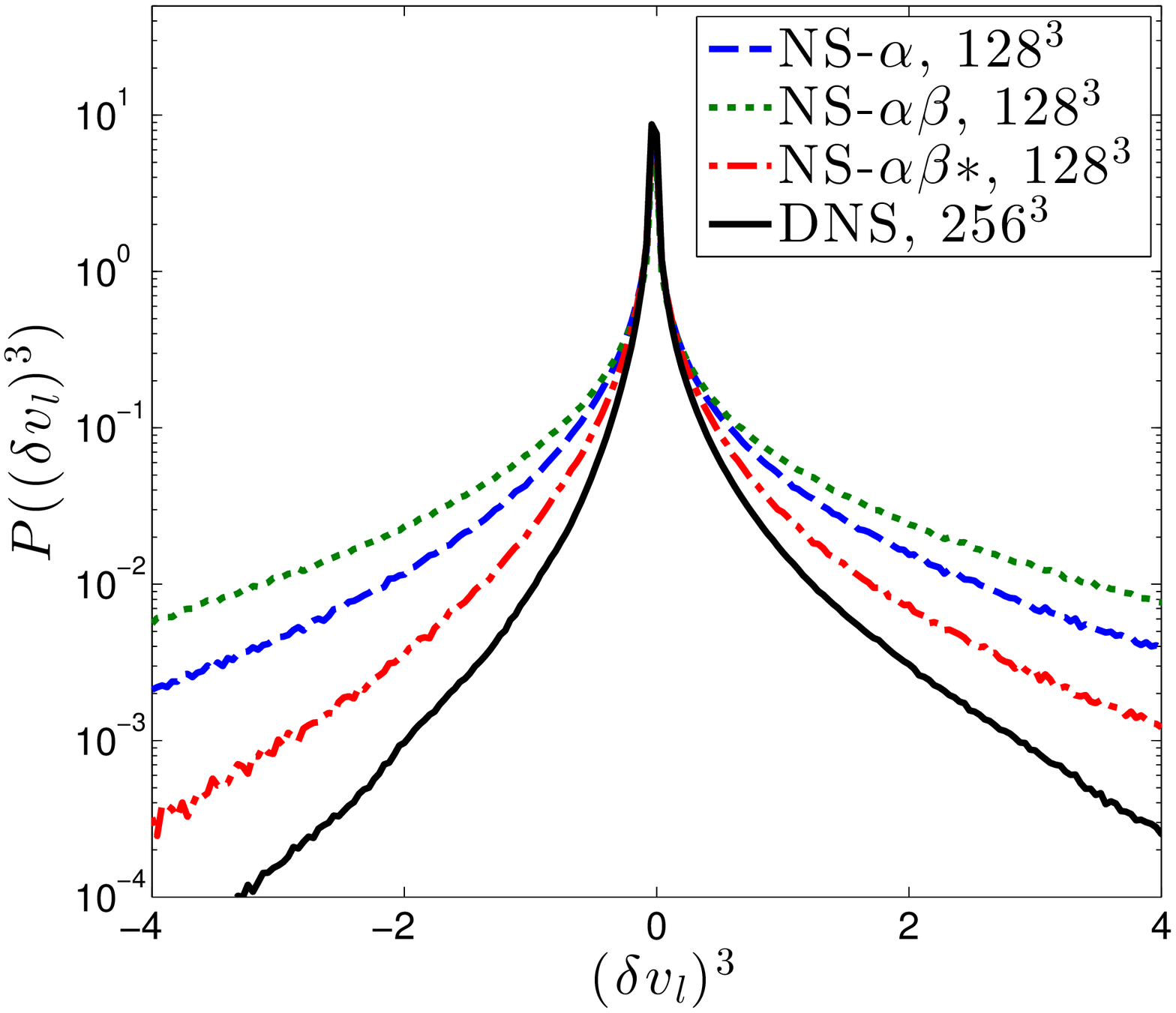,width=6.5cm, height=5.5cm}} 
 \put(230,-10){\epsfig{file=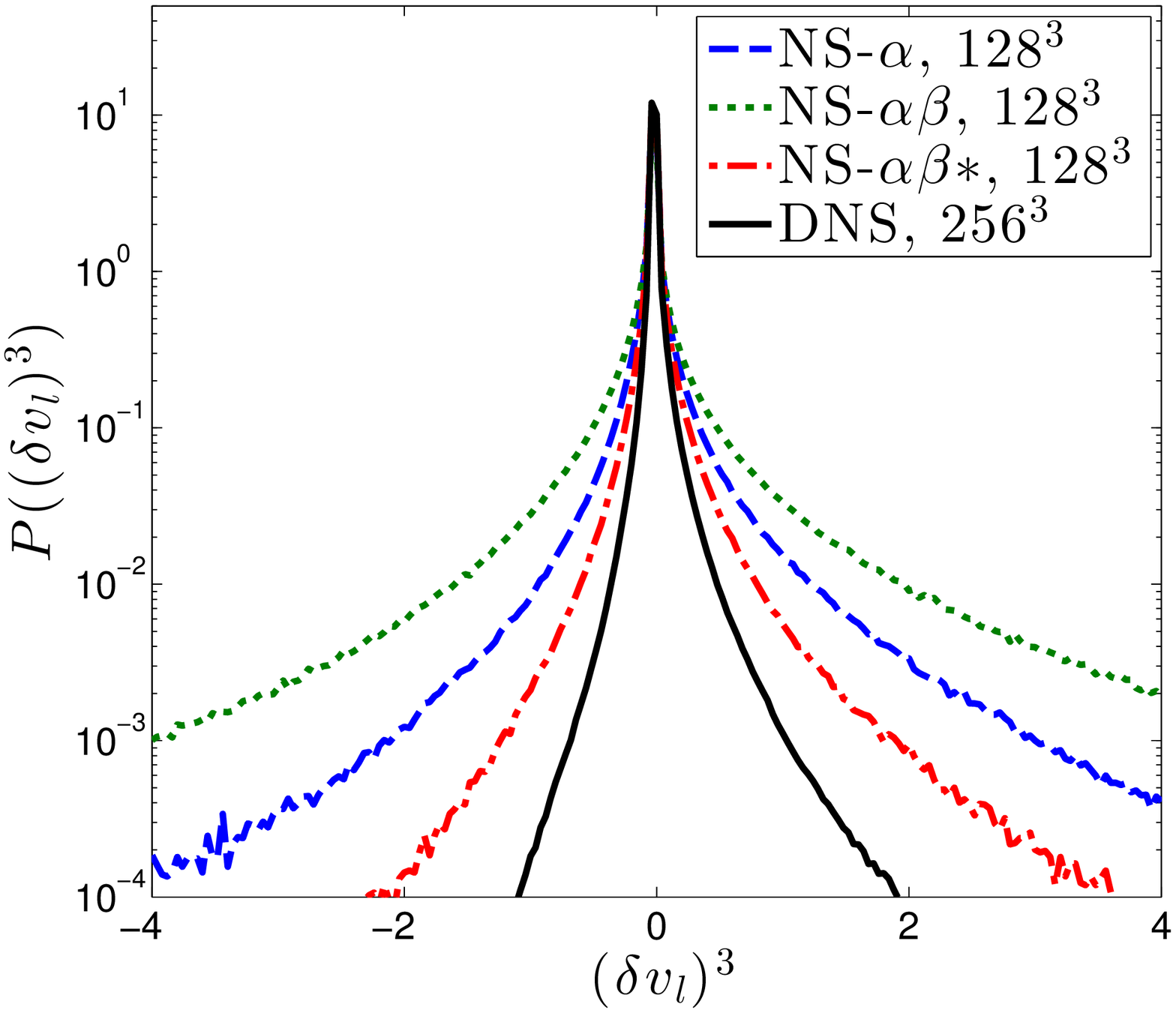,width=6.5cm, height=5.5cm}}

\put(50,450){\footnotesize(a) $r = 5/32$}
\put(50,290){\footnotesize(b) $r = 1/16$}
\put(50,130){\footnotesize(c) $r = 1/32$}

\put(260,450){\footnotesize(d) $r = 5/32$}
\put(260,290){\footnotesize(e) $r = 1/16$}
\put(260,130){\footnotesize(f) $r = 1/32$}

\end{picture}
\end{center}
 \caption{PDFs of the third-order moments of the filtered velocity increment $\delta u_l$ and the unfiltered velocity increment $\delta v_l$ for NS-$\alpha$ ($\alpha=1/8$) and NS-$\alpha\beta$ ($\alpha=1/8$, $\beta=1/12$) compared to NS-$\alpha\beta*$ ($\alpha=1/8, \beta=1/12$) at $128^3$ resolution and DNS at $256^3$ resolution with different normalized separation distances $r= 5/32$, $r= 1/16$, and $r= 1/32$.}
\label{PDF3}
\end{figure*}

\subsection{Two-point statistics}
\label{sec:PDF}

Two-point correlations are widely used to describe the spatial structure of turbulent velocity fields from a statistical viewpoint. Here, we focus on a particular instant $t$ of time and consider purely spatial correlations. For brevity and without loss of generality, we therefore suppress dependence on $t$. Structure functions are two-point correlations, linking the velocities $\bfv(\bfx)$ and $\bfv(\bfx+\bfr)$ at two different points $\bfx$ and $\bfx+\bfr$ in space separated by a vector $\bfr$ through the velocity increment
\begin{equation}
\delta \bfv(\bfx,\bfr) = \bfv(\bfx+\bfr) -  \bfv(\bfx).
\end{equation}
The projection of the velocity increment $\delta \bfv(\bfx,\bfr)$ onto the direction of $\bfr$ yields the scalar longitudinal velocity increment
\begin{equation}
\delta v_l(\bfx,\bfr) = \delta \bfv(\bfx,\bfr) \cdot \hat{\bfr}, \qquad \hat{\bfr}=\frac{\bfr}{r},
\end{equation}
where $r=|\bfr|$ is the separation distance with the dimension of length. For statistically homogeneous and isotropic turbulence, the statistics of $\delta v_l$ are independent of the direction $\hat{\bfr}$ but do depend on the modulus $r$. Hence, in homogeneous isotropic turbulence, the statistics of $\delta v_l$ can be evaluated for arbitrary directions $\hat{\bfr}$. We non-dimensionalize the separation distance $r$ with the domain size $L$ and keep the same symbols for convenience.
The dimensionless separation distance $r$ and the separation distance $N_r$ in mesh points are related through $r=N_r/N $, with $N$ being the number of grid points in a given direction.

Let $P(\delta v_l)$ denote the PDF of the unfiltered velocity increment $\delta v_l$; similarly, let $P((\delta v_l)^3)$ denote the PDF of the third moment of the unfiltered velocity increment $\delta v_l$. The PDFs $P(\delta u_l)$ and $P((\delta u_l)^3)$ of the filtered velocity increments are defined analogously. To obtain better statistics, PDFs are evaluated with velocity increments in all three Cartesian coordinate directions $\hat\bfx$, $\hat\bfy$, and $\hat\bfz$ and are averaged over ten data sets collected during approximately thirteen large eddy turnover times. We compute PDFs of the NS-$\alpha\beta$ model for three values, $r= 5/32$, $r= 1/16$, and $r= 1/32$, of the normalized separation distances, corresponding, respectively, to separation distances in mesh points of $N_r=40$, $N_r=16$, and $N_r=8$ on the $256^3$ grid. While large separation distances coincide with low wavenumbers in the inertial range, small separation distances coincide with large wavenumbers in the dissipation range.

Previous studies of DNS data by~Chen et al.~\cite{Chen1993} and Ishihara et al.~\cite{Ishihara2009} showed that the normalized PDFs at large separation distance are close to Gaussian. Furthermore,  the tails of the PDFs at smaller separation distances are expected to exhibit nearly exponential behavior. 
This behavior is evident in the PDFs for $256^3$ DNS results shown in Figure~\ref{PDF1n} along with the normalized PDFs of the velocity increments of the NS-$\alpha\beta*$ model at $128^3$ resolution and the PDFs of the velocity increment for the unfiltered and filtered velocity fields of the NS-$\alpha$ and NS-$\alpha\beta$ models at $128^3$ resolution. 
Consistent with the discussion of structure functions provided in Section~\ref{sec:struct}, the PDFs are normalized with the square root of the second moment $\langle (\delta u_l)^2\rangle^{1/2}$ of the velocity increment at the corresponding separation distance $r$. 

The normalized PDFs of the NS-$\alpha$ model, the NS-$\alpha\beta$ model, and the NS-$\alpha\beta*$ model are indistinguishable from those of the DNS results in a neighbourhood of what appears to be a shared maximum. 
However, the tails of the PDFs of the unfiltered and filtered velocity increment deviate from those of the corresponding DNS results. 
The NS-$\alpha\beta*$ model is found to exhibit broader tails of the PDFs than the corresponding DNS results at all separation distances. In accord with the preceding discussion, the NS-$\alpha\beta*$ model and the DNS results possess only one velocity field and thus the same PDFs of the DNS results and NS-$\alpha\beta*$ results are shown in panels (a)--(c) and (d)--(f) of Figure~\ref{PDF1n} and all the following figures.
In general, the normalized PDFs of the filtered velocity increments of the NS-$\alpha$ and the NS-$\alpha\beta$ model are narrower than the corresponding PDFs of the DNS velocity increments, as shown in panels (a)--(c) of Figure~\ref{PDF1n}. On the contrary, the PDFs of the unfiltered velocity increments of the NS-$\alpha$ and NS-$\alpha\beta$ model have broader tails than the PDF from the DNS results, as shown in panels (d)--(f) of Figure~\ref{PDF1n}. This effect becomes more prominent with decreasing separation distance $r$. Since the normalized PDFs of the filtered and unfiltered velocity fields nearly coincide with the DNS results, the filtering operation has no influence on the large scales; however, the difference between the PDFs obtained from the filtered and unfiltered velocity fields increases with decreasing separation distance $r$ or, equivalently, with decreasing scale. Since small separation distances correspond to the dissipation range, we confirm the view that both regularization models alter the small scale properties of the flow, while providing a good approximation for the large scale properties of the flow. For all three separation distances, the tails of the normalized PDFs of the NS-$\alpha\beta$ model are slightly closer to the DNS results than those of the NS-$\alpha$ model. However, in both regularization models, the tails of the normalized PDFs of the unfiltered velocity field exhibit exponential behavior, even at intermediate separation distances. This result indicates that the unfiltered velocity fields of both regularization models exhibit unphysical statistics, which improve to become more realistic through the filtering operation.

Besides the discussion of the normalized PDFs in Figure~\ref{PDF1n}, which gives an idea of the relative shape of the PDFs, the ``raw" PDFs, shown in Figure~\ref{PDF1} are useful to evaluate and compare the actual values of the filtered and unfiltered longitudinal velocity increments and to relate the real space statistics to the statistics in wavenumber space. 
Notice that the PDFs of the unfiltered longitudinal velocity increments of the NS-$\alpha$ and NS-$\alpha\beta$ models and the PDFs of the velocity increments of the NS-$\alpha\beta*$ model are, for all three choices of separation distances, significantly broader than the PDF of the velocity increments of the $256^3$ DNS results. The NS-$\alpha\beta$ model gives higher probabilities of larger unfiltered velocity increments than the NS-$\alpha$ model and the NS-$\alpha\beta*$ model at all separation distances, but the difference between the PDFs of NS-$\alpha$ and NS-$\alpha\beta$ models increases with decreasing separation distances. 
This observation seems consistent with the notion that $\beta$ is associated with the dissipation range. The PDFs of the unfiltered longitudinal velocity increments of the NS-$\alpha$ model are closer to the PDFs obtained from the $256^3$ DNS results than are the PDFs of the unfiltered longitudinal velocity increments of the NS-$\alpha\beta$ model. As already mentioned in the discussion of the normalized PDFs of the unfiltered longitudinal velocity increments, the unfiltered velocity fields of the NS-$\alpha$ and NS-$\alpha\beta$ model do not appear to possess physically good statistical properties when compared to the statistical properties of DNS results at higher resolution.

The PDFs of the filtered velocity increments are shown in panels (a)--(c) of Figure~\ref{PDF1} along with the PDFs of the velocity increments of the NS-$\alpha\beta*$ model and the DNS results. 
For each of the separation distances considered, the PDFs of the filtered velocity increments of the NS-$\alpha$ and NS-$\alpha\beta$ model are very similar and provide good approximations to the PDFs obtained from the DNS results. The PDFs of the filtered velocity increments of the NS-$\alpha\beta$ model tend to be slightly closer to the PDFs obtained from the DNS results. At large separation distances, the PDFs of the filtered velocity increments of the NS-$\alpha$ and NS-$\alpha\beta$ model closely approximate the PDF of the DNS results and the difference between PDFs of the NS-$\alpha$ and NS-$\alpha\beta$ model and the DNS results increases with decreasing separation distance $r$. This confirms the view that large scale features of the flow are well approximated by the NS-$\alpha$ and NS-$\alpha\beta$ model while smaller scales flow features are lost through filtering. An optimal choice of parameters $\alpha$ and $\beta$ was not attempted in the current study and might well yield further improvements. 
The PDFs of the velocity increments of the NS-$\alpha\beta*$ model provides a further indicator that the modified viscus term of the NS-$\alpha\beta$ model allows for more small scale activity, leading to broader PDFs of the velocity increments. However, the PDF statistics of the NS-$\alpha\beta*$ model confirm that the NS-$\alpha\beta*$ model does not provide statistics suitable to qualify it as a regularization model, as already diagnosed in the discussion of the energy spectra.

Drawing a parallel between statistics in wavenumber space presented in the form of the energy spectra in Figure~\ref{espE} and \ref{espEuv} and statistics in physical space leads to the following explanation. The natural energy norm for the NS-$\alpha$ and NS-$\alpha\beta$ model involves the inner product $\bfv \cdot \bfu$. The unfiltered velocity field $\bfv$ of the NS-$\alpha\beta$ model is observed to possess broader PDFs of the $\delta v_l$ velocity increments corresponding to higher probabilities of larger velocity increments, especially at intermdiate and small separation distances. Consequently, it appears logical that, at higher wavenumbers, the energy spectra of the NS-$\alpha\beta$ model are closer to the energy spectra of DNS than the energy spectra of the NS-$\alpha$ model. That is, through the availability of what might be viewed as an additional tuning parameter, namely $\beta<\alpha$, the NS-$\alpha\beta$ model allows for more small scale motion than the NS-$\alpha$ model. 

The PDFs $P((\delta v_l)^3)$ and $P((\delta u_l)^3)$ of the third-order moments of the unfiltered and filtered longitudinal velocity increments are shown in Figure~\ref{PDF3}. They show the same qualitative behavior as the PDFs of the unfiltered and filtered longitudinal velocity increments, but the mentioned tendencies become more prominent in the third-order moments, representing a higher-order statistics of the velocity field. While the PDFs of the third-order moment of the unfiltered velocity field of the NS-$\alpha\beta$ model and the NS-$\alpha$ model are very similar at large separation distances within the inertial range, the differences between the two models increase with decreasing separation distance $r$, confirming the view that the parameter $\beta$ is responsible for the properties of the model at small scales. From the PDFs of the third-order moments of the filtered longitudinal velocity increments at intermediate separation distance, we observe that the results from the  NS-$\alpha\beta$ model are closer to the results obtained from DNS than the results from the NS-$\alpha$ model over a wide range of the PDF.

\subsection{Structure functions}
\label{sec:struct}
\begin{figure*}[!t]
\begin{center}
\begin{picture}(450,270)

 \put(0,  140){\epsfig{file=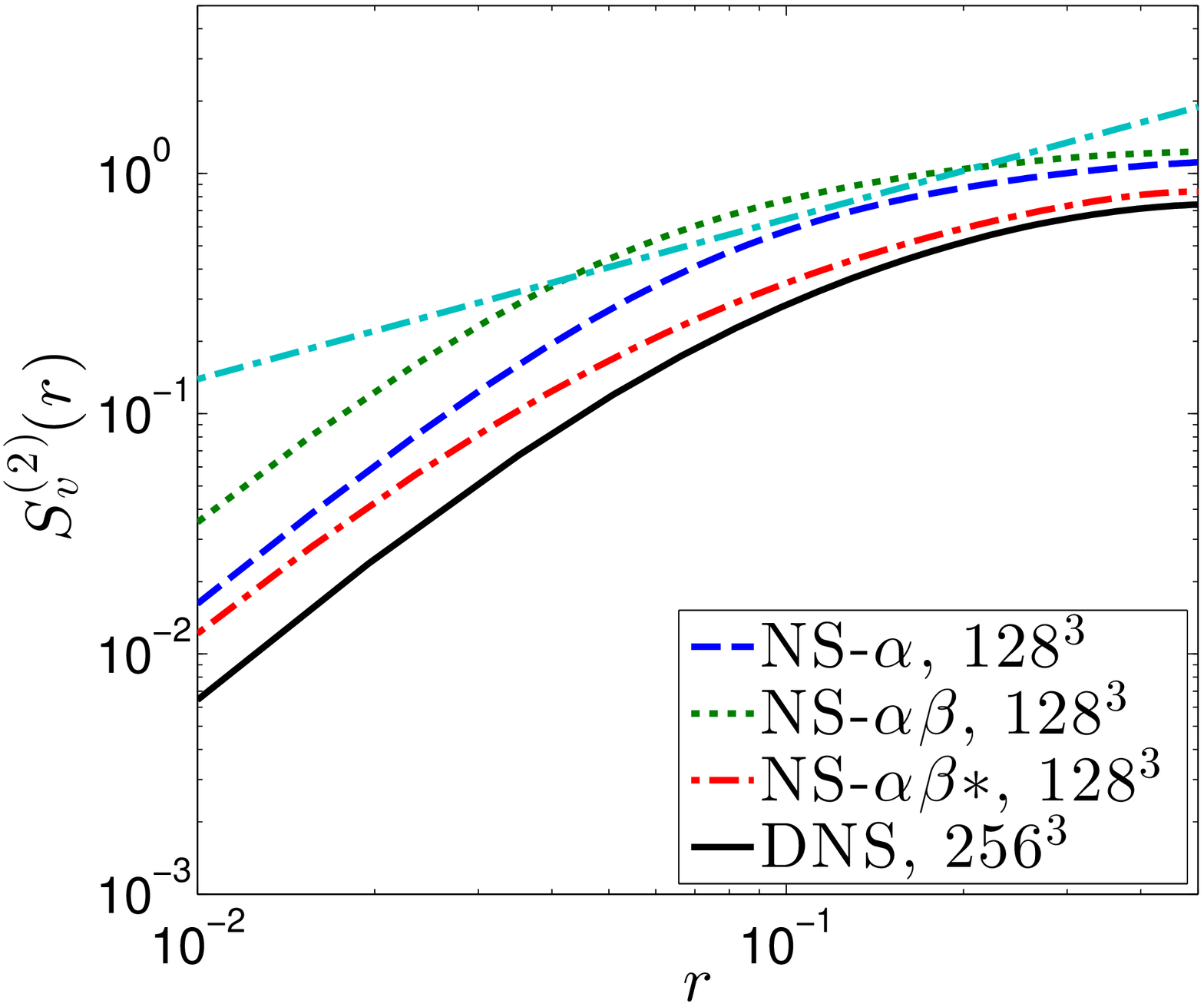,width=5cm,height=4.2cm}}
 \put(145,140){\epsfig{file=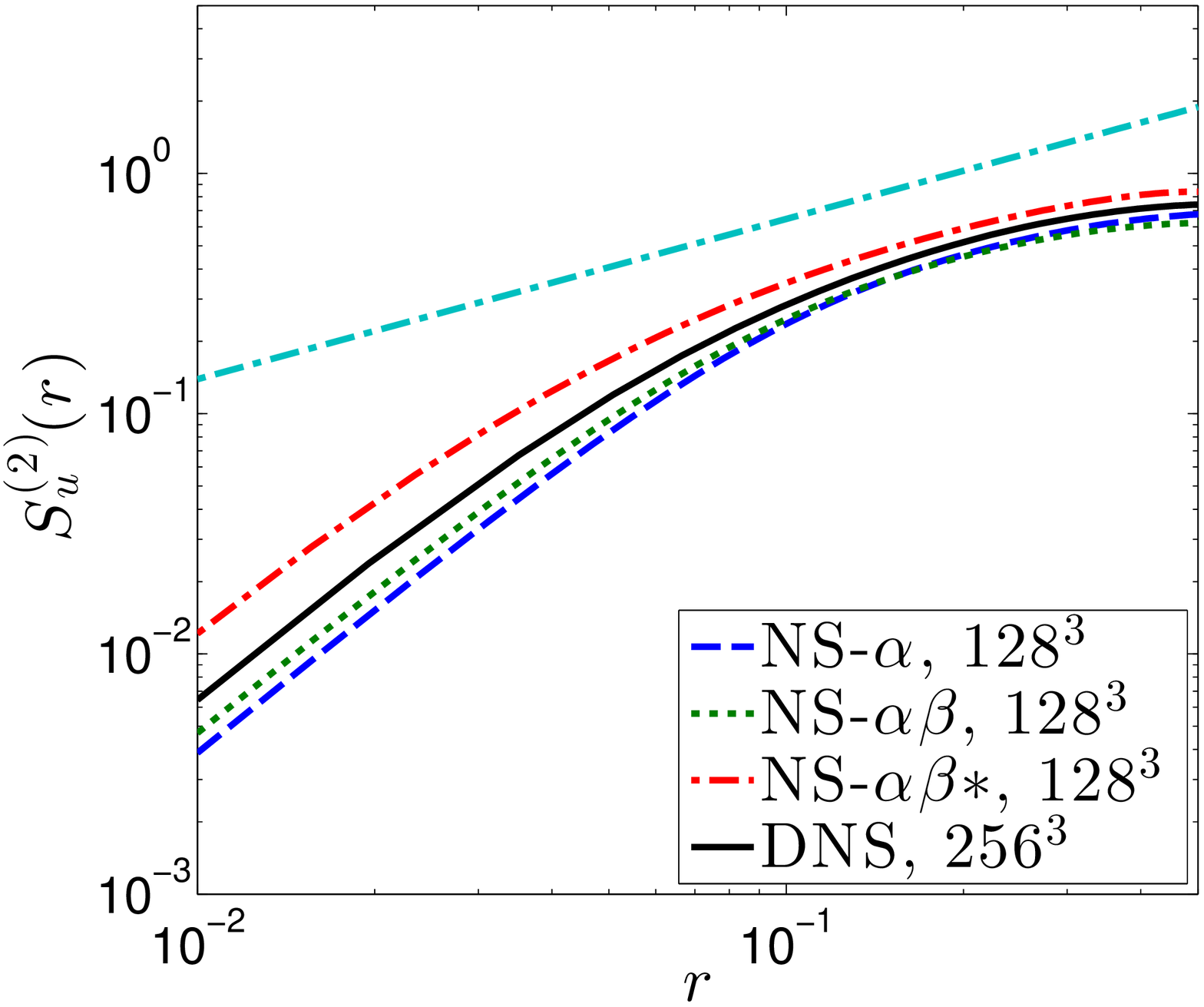,width=5cm,height=4.2cm}}
 \put(290,140){\epsfig{file=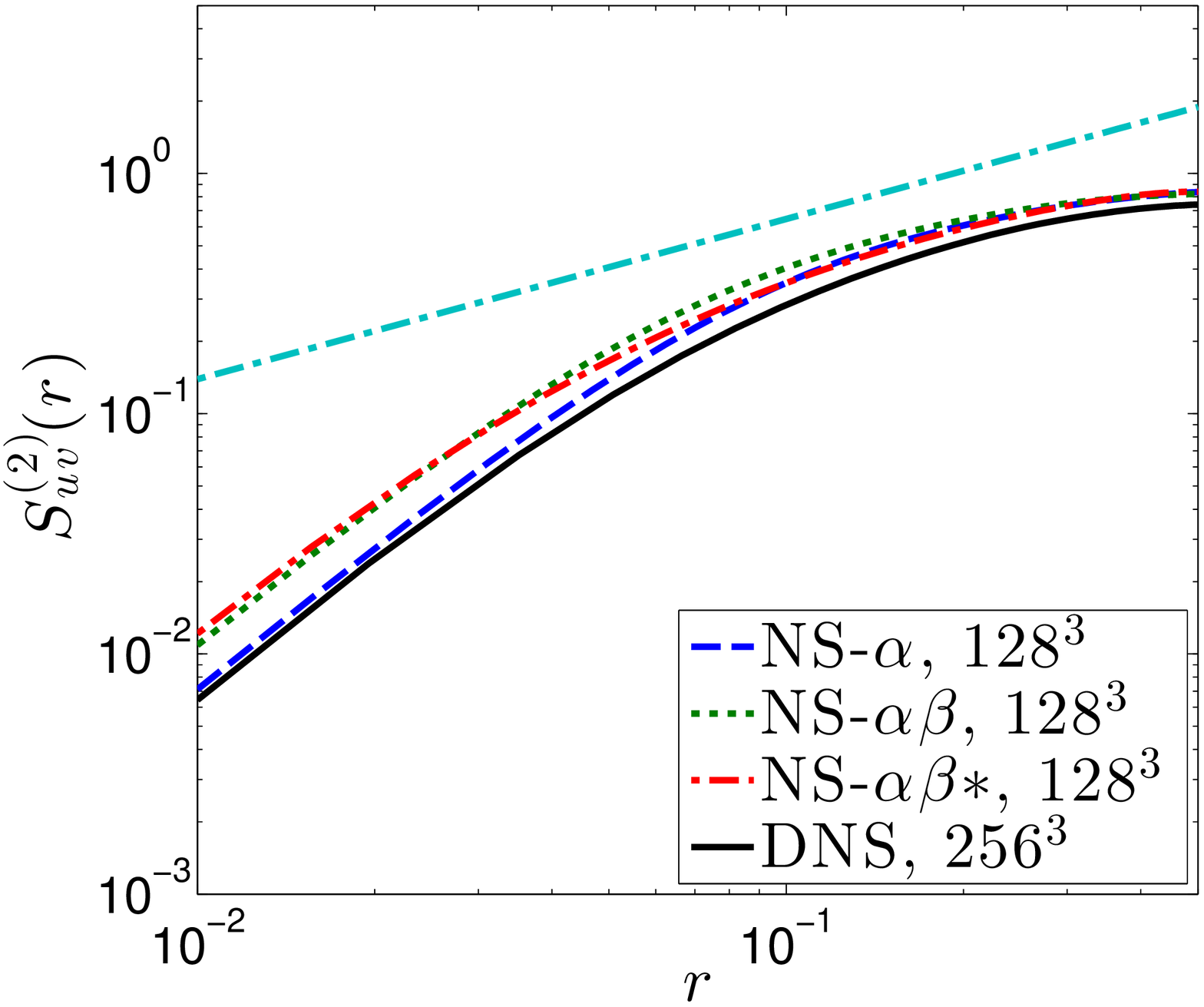,width=5cm,height=4.2cm}}

 \put(0,  0){\epsfig{file=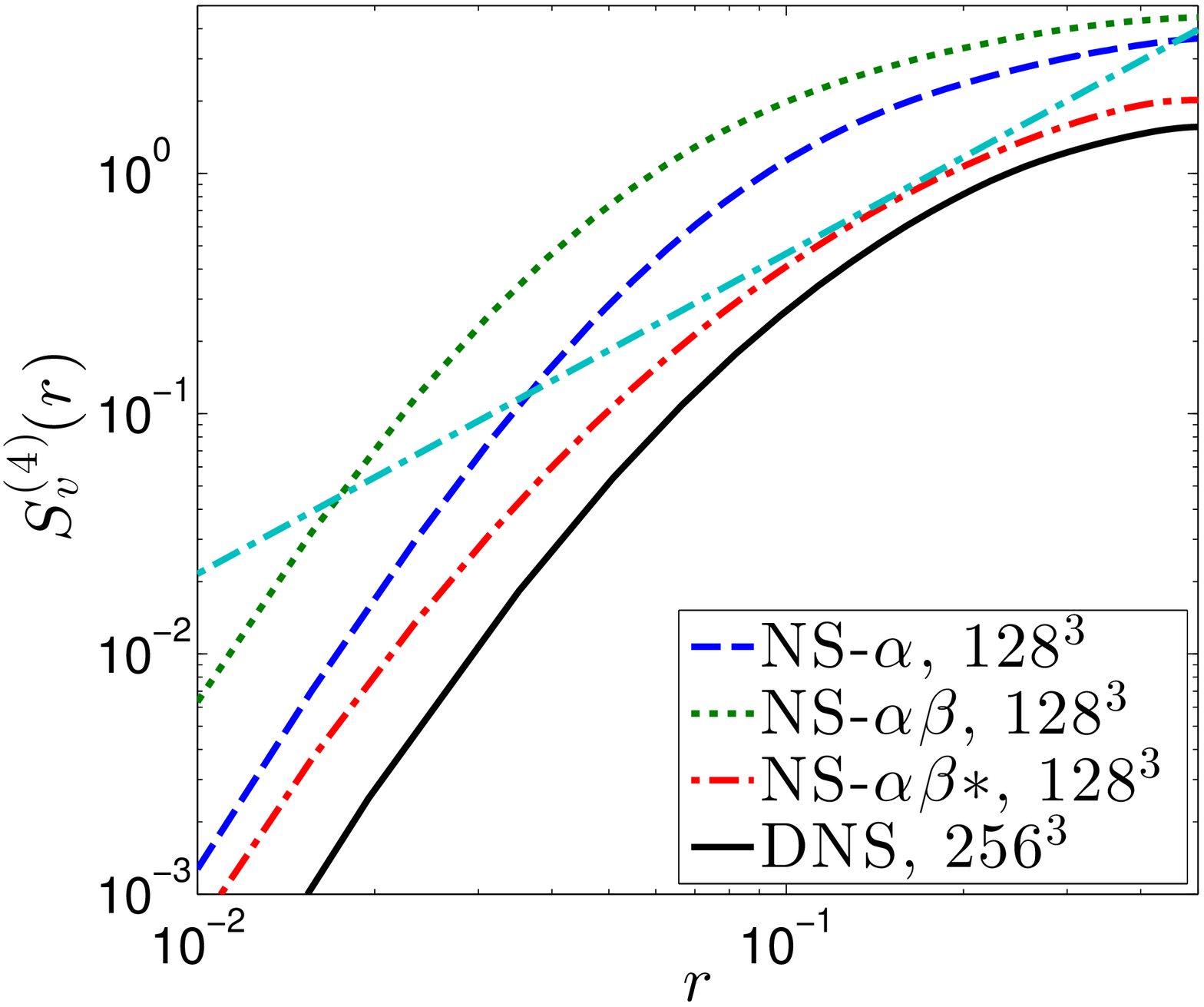,width=5cm,height=4.2cm}}
 \put(145,0){\epsfig{file=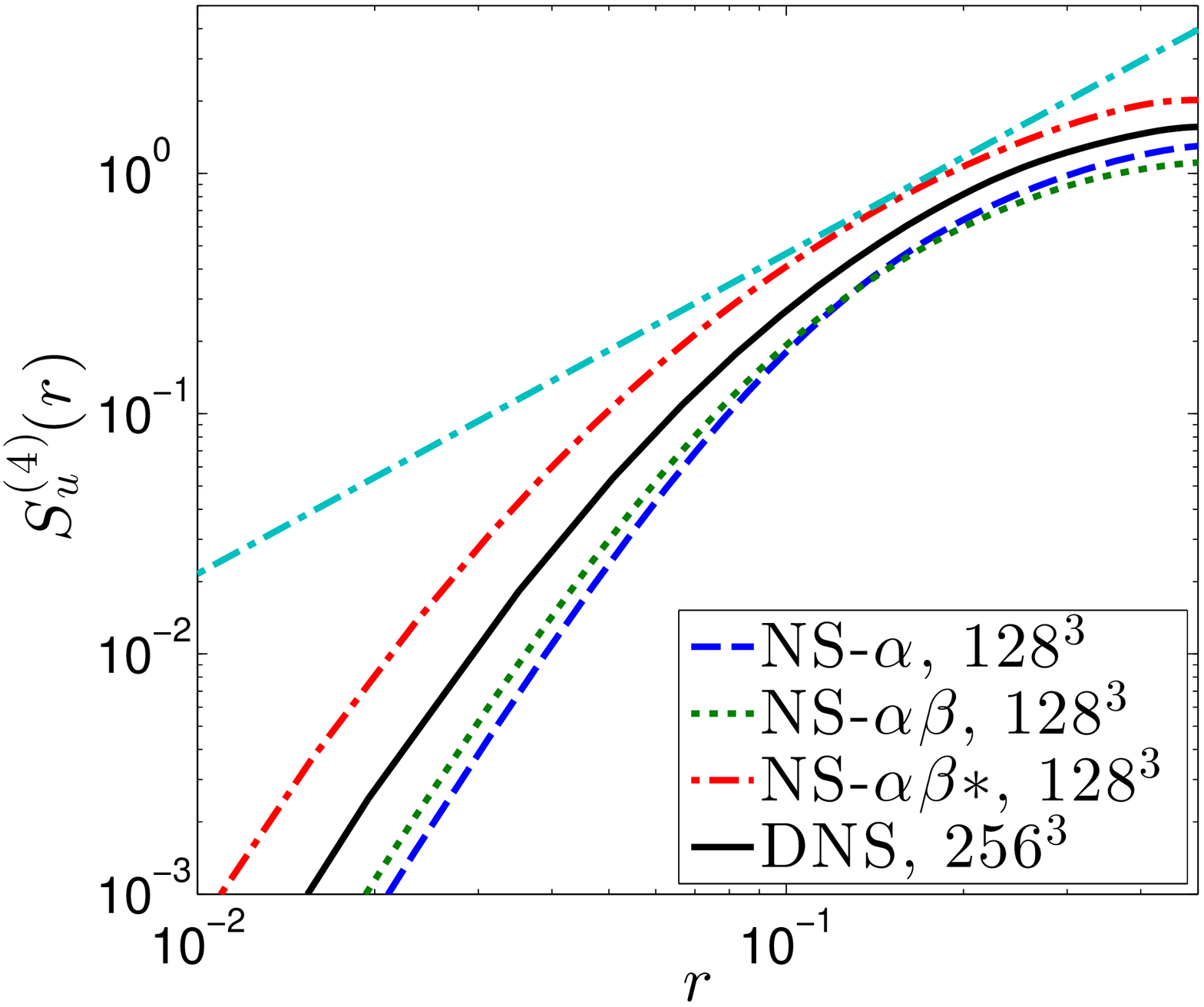,width=5cm,height=4.2cm}}
 \put(290,0){\epsfig{file=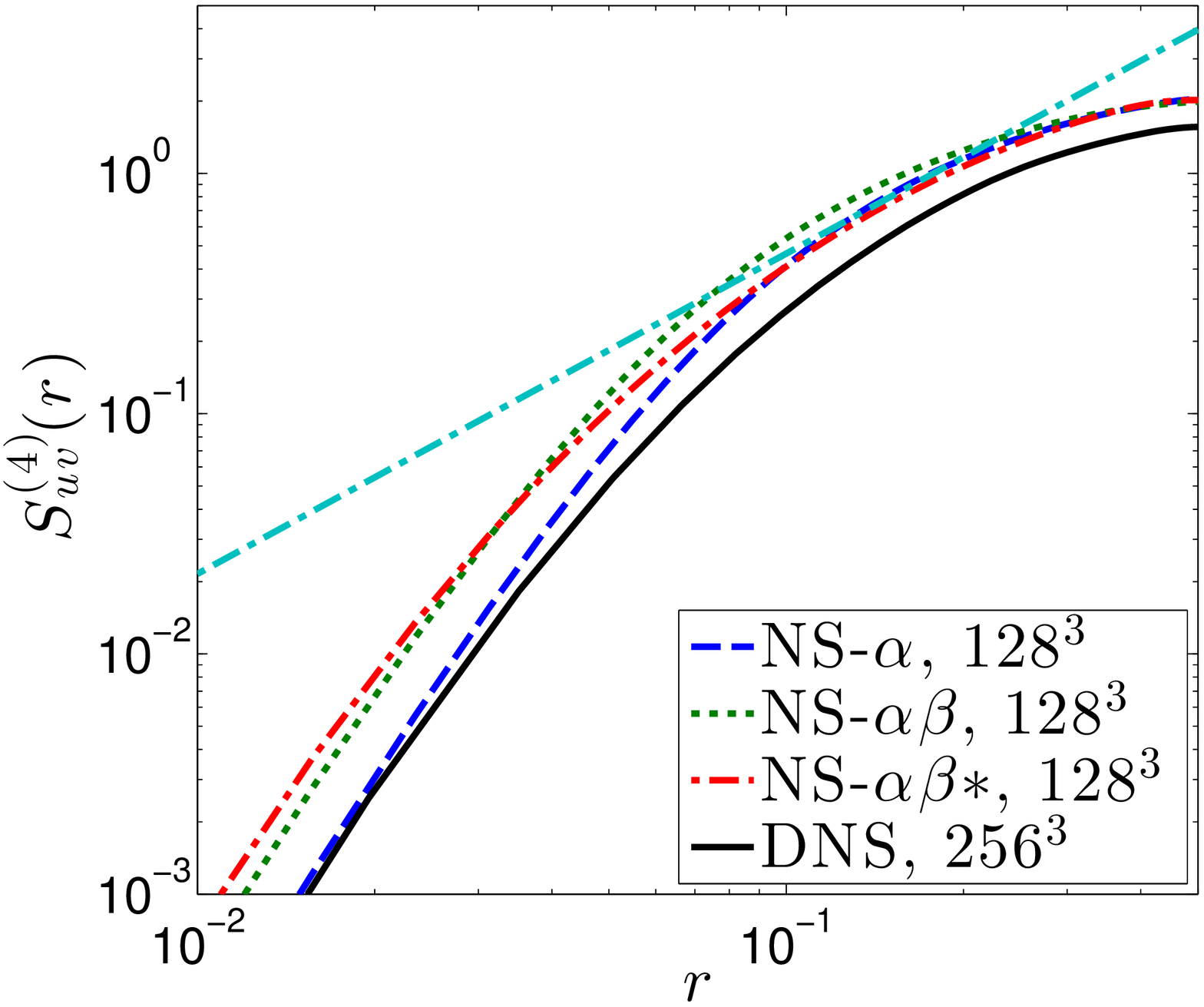,width=5cm,height=4.2cm}}
 \put(72.5,125){(a)}
 \put(220,125){(b)}
 \put(362.5,125){(c)}
 \put(30,225){$r^{2/3}$}
 \put(177.5,225){$r^{2/3}$}
 \put(325,225){$r^{2/3}$}
 \put(72.5,-15){(d)}
 \put(220,-15){(e)}
 \put(362.5,-15){(f)}
  \put(30,65){$r^{4/3}$}
 \put(177.5,65){$r^{4/3}$}
 \put(325,65){$r^{4/3}$}

 \end{picture}
 \end{center}
 \caption{Second- and fourth-order structure functions of NS-$\alpha$ ($\alpha=1/8$) and NS-$\alpha\beta$ ($\alpha=1/8$, $\beta=1/12$) compared to NS-$\alpha\beta*$ ($\alpha=1/8, \beta=1/12$) at $128^3$ resolution and DNS results at $256^3$ resolution and the K41 scaling prediction; (a)--(c) second-order structure functions of unfiltered, filtered, and mixed velocity increments (left to right); (d)--(f) fourth-order structure functions of unfiltered, filtered, and mixed velocity increments (left to right).}
\label{struct24}
\end{figure*}

The scalar, longitudinal structure function of order $n$ is defined as
\begin{equation}
S^{(n)}(r)= \left< \delta v_l(r)^n \right>,
\end{equation}
where $\left<\cdot\right>$ denotes a suitable ensemble average. The ``four-fifths" law as part of Kolmogorov's famous theory (see, for example Frisch~\cite{Frisch1996}), dating from 1941 (and denoted by K41 hereafter), relates the third-order velocity structure function in the inertial range of homogeneous isotropic turbulence with the dissipation rate $\epsilon$ and predicts a scaling 
\begin{equation}
S^{(3)}(r)= \left\langle  \delta v_l(r)^3 \right\rangle  = -\frac{4}{5}\epsilon r
\end{equation}
proportional to the separation distance $r$. More generally, Kolmogorov concluded that the structure function of order $n$ scales as
\begin{equation}\label{eq:Sn}
S^{(n)}(r)= \left\langle  \delta v_l(r)^n \right\rangle  = -C_n(\epsilon r)^{n/3},
\end{equation}
where the coefficient $C_n$ is a universal constant. For higher-order structure functions, scaling exponents have been found to deviate from the K41 scaling prediction due to intermittency, as discussed by Frisch~\cite{Frisch1996} and others.
We discuss structure functions 
\begin{equation}
S_v^{(n)}(r)= \left\langle  \delta v_l(r)^n \right\rangle
\end{equation}
of the unfiltered velocity field, structure functions 
\begin{equation}
S_u^{(n)}(r)= \left\langle  \delta u_l(r)^n \right\rangle
\end{equation}
of the filtered velocity field, and mixed structure functions
\begin{equation}
S_{uv}^{(n)}(r)= \left\langle  (\delta u_l(r)\delta v_l(r))^{n/2} \right\rangle.
\end{equation}
In detail, the procedure we use to compute the structure functions in physical space is as follows. For a velocity field at a fixed time $t$, we choose $\hat{\bfr}$ to coincide with one of the Cartesian unit directions $\hat{\bfx}$, $\hat{\bfy}$, or $\hat{\bfz}$. For each separation parameter $r$, at each discrete grid point $\bfx_d$ in the computational domain, we evaluate the velocity increments between the neighbouring points located at $\bfx_d+  \frac{r}{2}\hat{\bfx}$ and  $\bfx_d- \frac{r}{2}\hat{\bfx}$, respectively. The velocity increments, or their higher-order moments are averaged over all points in the computational domain and determine one value $S^{(n)}(r)$ of the structure function. This procedure is repeated for all separation distances $r$ in all three Cartesian directions. That is, we compute structure functions in the three Cartesian coordinate directions and average them to obtain more representative statistical results. This procedure is repeated for ten data sets collected during approximately thirteen large eddy turnover times and the results are averaged.

Figure~\ref{struct24} shows the second- and fourth-order unfiltered, filtered, and mixed structure functions of the NS-$\alpha$ ($\alpha=1/8$) and NS-$\alpha\beta$ ($\alpha=1/8$, $\beta=1/12$) models at $128^3$ resolution compared to structure functions of the NS-$\alpha\beta*$ model ($\alpha=1/8$, $\beta=1/12$) at $128^3$ resolution, DNS at $256^3$ resolution, and the K41 scaling prediction. 
In the second- and fourth-order structure functions of the $256^3$ DNS, we identify a short section around $r \approx 0.15$ where they scale with the corresponding K41 prediction. Notice that $r \approx 0.15$ corresponds approximately to the same regime as $r=5/32$, the separation distance for which we also computed PDFs of the longitudinal velocity increments presented and discussed above. This confirms that $r=5/32$ corresponds to a separation distance within the inertial range. 
The second- and fourth-order structure functions of the NS-$\alpha\beta*$ model show higher values than the corresponding structure functions of the $256^3$ DNS data. This observation is in agreement with the previously discussed energy spectra of the NS-$\alpha\beta*$ model and confirms that the modified viscous term in the NS-$\alpha\beta*$ model and the NS-$\alpha\beta$ model allows for more small scale activity compensating for a possible overdamping through the NS-$\alpha$ SGS stress term.
The second- and fourth-order structure functions of the unfiltered velocity field of the NS-$\alpha$ and NS-$\alpha\beta$ model exhibit significantly higher values than the $256^3$ DNS along the full range of separation distances. The NS-$\alpha\beta$ model possesses higher moments of the unfiltered longitudinal velocity increments than the NS-$\alpha$ model. In the inertial range, the NS-$\alpha$ and NS-$\alpha\beta$ structure functions of the unfiltered velocity  both appear to scale with a smaller exponent than the corresponding K41 scaling prediction. The second- and fourth-order filtered velocity structure functions of the NS-$\alpha$ and NS-$\alpha\beta$ model closely approximate the respective structure functions of the $256^3$ DNS results. Near $r\approx 0.15$, the filtered structure functions of the NS-$\alpha$ and NS-$\alpha\beta$ model coincide in the inertial range; further, for smaller separation distances $r$, the NS-$\alpha\beta$ model exhibits higher values than than the NS-$\alpha$ model and, thus, in the dissipation range provides a better approximation to the $256^3$ DNS results than the NS-$\alpha$ model. This behavior points to the presence of increased motion at smaller scales in the results of the NS-$\alpha\beta$ model compared to the NS-$\alpha$ model and is consistent with the behavior of the PDFs discussed in Section~\ref{sec:PDF}. The mixed velocity structure functions of the NS-$\alpha$ and NS-$\alpha\beta$ model both show higher values than the corresponding $256^3$ DNS. For separation distances $r\gtrsim 0.15$ the values of NS-$\alpha$ and NS-$\alpha\beta$ model coincide for both, second- and fourth-order mixed structure functions. At smaller separation distances, the NS-$\alpha\beta$ model displays higher values than the NS-$\alpha$ model and the mixed structure functions of the NS-$\alpha$ model provide a better approximation to the $256^3$ DNS results. 
\subsection{Flatness of the velocity increments}
\begin{figure*}[!t]
\begin{center}
\begin{picture}(420,200)
\put(0,  0){\epsfig{file=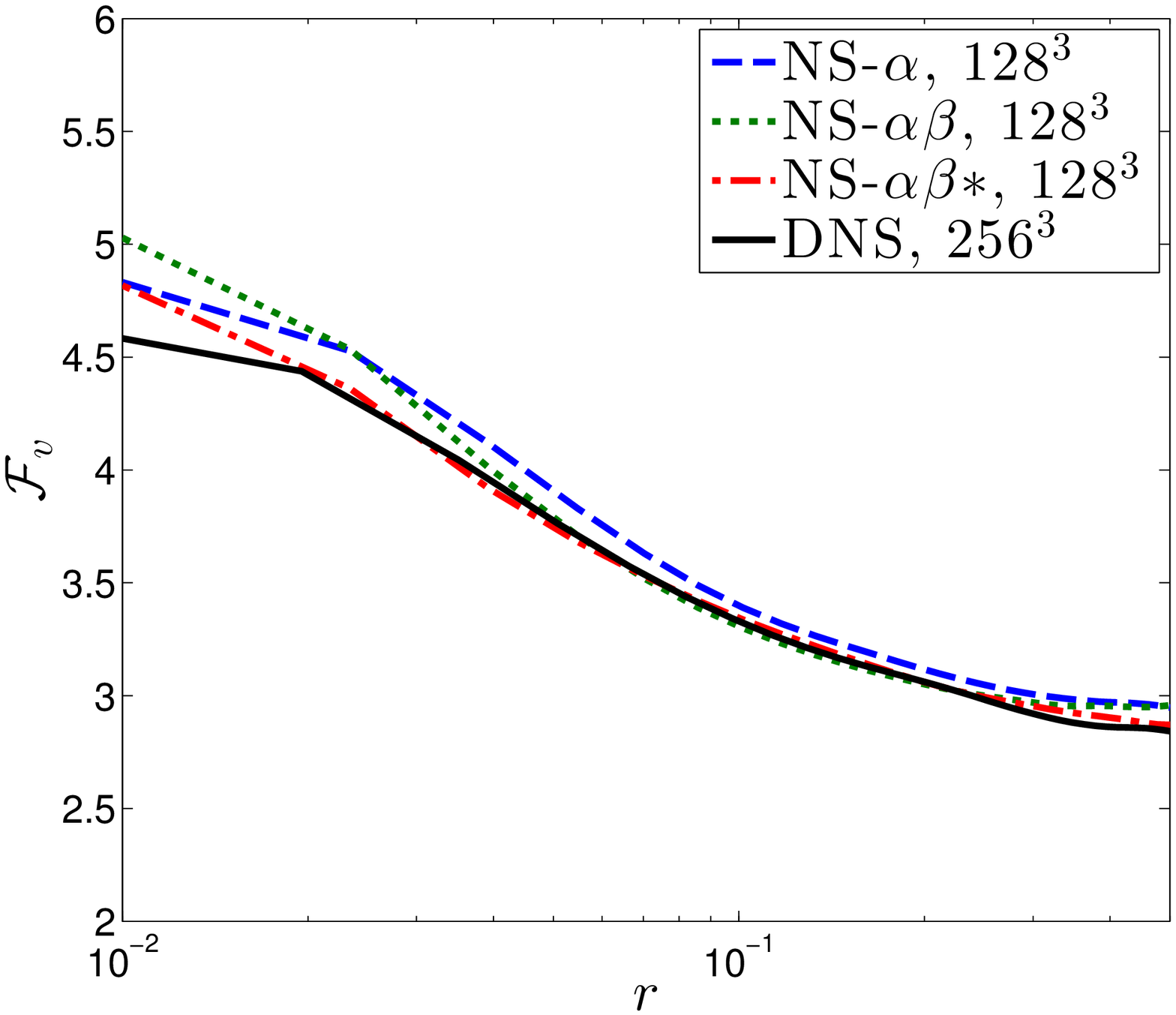, width=7.5cm}}
\put(220,0){\epsfig{file=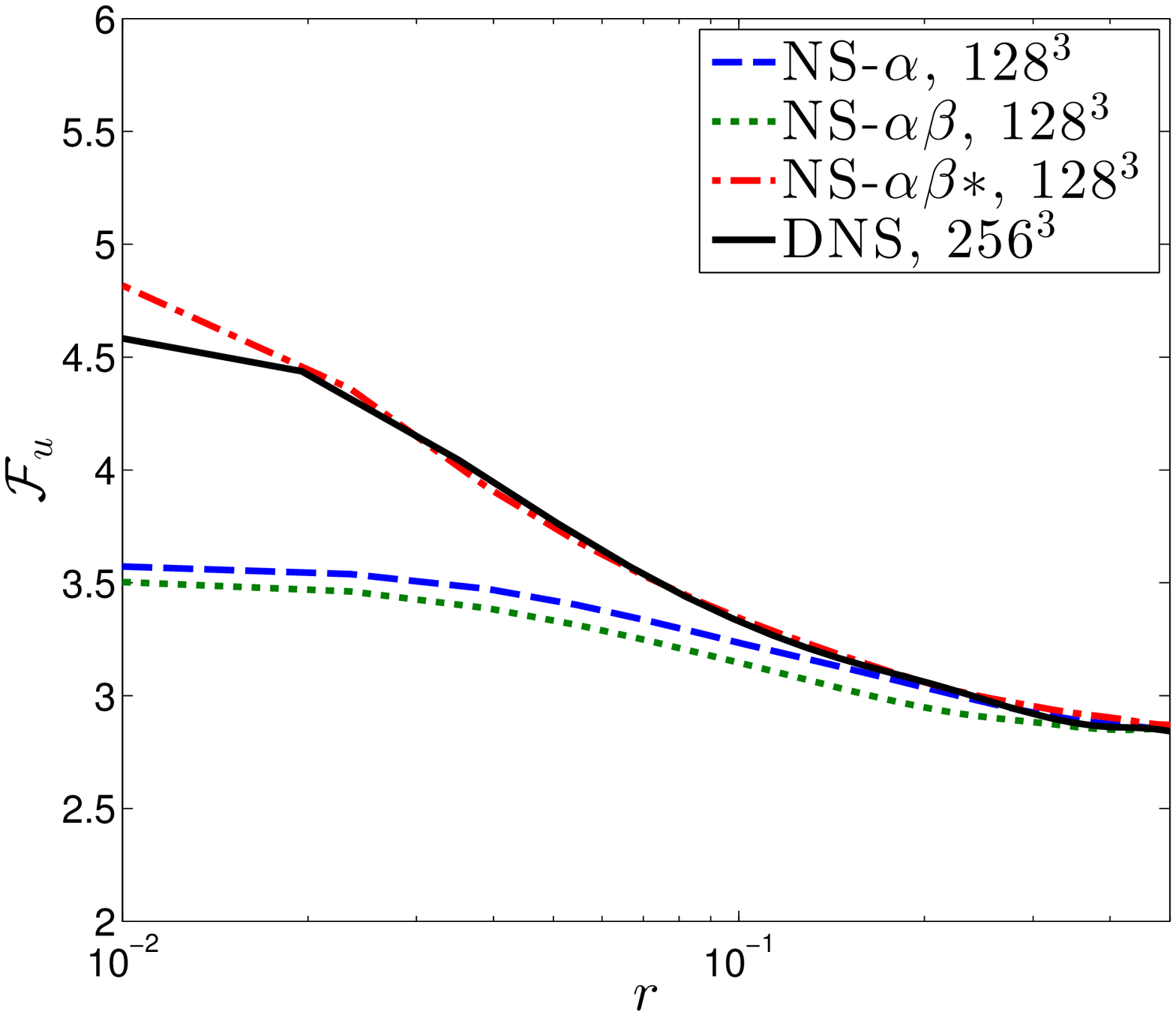, width=7.5cm}}
 \put(100,-15){(a)}
 \put(330,-15){(b)}
 \end{picture}
 \end{center}
 \caption{Flatness factors ${\cal{F}}_v$ and ${\cal{F}}_u$ of the unfiltered and filtered velocity structure function of NS-$\alpha$ ($\alpha=1/8$) and NS-$\alpha\beta$ ($\alpha=1/8$, $\beta=1/12$) compared to NS-$\alpha\beta*$ ($\alpha=1/8, \beta=1/12$) at $128^3$ resolution and $256^3$ DNS results: (a) ${\cal{F}}_v$, (b) ${\cal{F}}_u$.}
\label{flat}
\end{figure*}

Central to the K41 theory is the assumption that the turbulent scales of motion are self-similar. This implies that the statistical features of turbulent flow should be independent of spatial scale. This self-similarity can be measured by the non-dimensional ratio of the fourth-order velocity structure function and the squared second-order velocity structure function. The flatnesses factors ${\cal{F}}_v$ and ${\cal{F}}_u$ of the unfiltered and filtered velocity increments are defined as
\begin{equation}
{\cal{F}}_v(r) = \frac{\langle \delta v_l^4 \rangle}{\langle \delta v_l^2\rangle^2},\qquad \quad  {\cal{F}}_u(r) = \frac{\langle \delta u_l^4 \rangle}{\langle \delta u_l^2\rangle^2},
\end{equation}
respectively.
If a flatness factor is flat or constant over all separation distances, the velocity distribution is self-similar. Otherwise, it is not self-similar, but instead it is characterized by intermittency.

As observed in studies of DNS reported by Cao et al.~\cite{Cao1996}, the flatness factor of NS turbulence is not constant as a function of the separation distance, due to the intermittent properties of the NS velocity fields. This observation is confirmed in our flatness results of our DNS results, shown in Figure~\ref{flat}. 
The NS-$\alpha\beta*$ model exhibits flatness factors almost identical to the flatness factors of the $256^3$ DNS results over a wide range of large and intermediate separation distances. At small separation distances, the NS-$\alpha\beta*$ model shows flatness factors slightly higher than those of the $256^3$ DNS results. Consequently, the modified viscous term in the NS-$\alpha\beta*$ model (and the NS-$\alpha\beta$ model) appears to influence the flatness statistics only at the smallest scales.
Chen et al.~\cite{Chen1999b} reported flatness factors for the filtered velocity field of the NS-$\alpha$ model and found that they are systematically lower than those of corresponding DNS velocity fields. This indicates that the filtered velocity field of the NS-$\alpha$ model is less intermittent than the velocity field obtained from DNS results. In our simulation results, the flatness of the filtered velocity field of the NS-$\alpha\beta$ model is found to be very close to the flatness of the NS-$\alpha$ model, but slightly lower throughout all separation distances, as shown in panel (b) of Figure~\ref{flat}. Apart from the flatness factors of the filtered velocity field, here, we also discuss the flatness factors of the unfiltered velocity fields of both NS-$\alpha$ and NS-$\alpha\beta$ model. Interestingly, the unfiltered velocity increments for both NS-$\alpha$ and NS-$\alpha\beta$ models exhibit slightly higher flatness factors than the velocity field from corresponding DNS results, as displayed in panel (a) of Figure~\ref{flat}. This result indicates that the unfiltered velocity fields of those two regularization models are more intermittent than the DNS results at higher resolution. However, the flatness factor of the unfiltered velocity field of the NS-$\alpha\beta$ model closely approximates the flatness factor of the DNS results over a wide range of intermediate separation distances, corresponding to the inertial range. That the unfiltered velocity field shows even higher flatness factors, and thus is more intermittent than DNS results, confirms our view that the statistical properties of the unfiltered velocity field are not sufficiently good to be considered a physically viable approximation to the velocity fields obtained from DNS results.

\section{Summary and conclusions}

We presented and discussed Fourier and real space statistics of the NS-$\alpha\beta$ model in homogeneous isotropic turbulence. 
Findings were compared to the limit case ($\beta=\alpha$) of the NS-$\alpha$ model, the limit case of the NS-$\alpha\beta*$ model ($\bftau_\alpha = \zed$), and DNS results at higher resolution. 
We confirmed the result of Kim et al.~\cite{Kim2009} showing that the natural energy spectrum of the NS-$\alpha\beta$ can closely approximate DNS results at higher resolution, given a suitable choice of the parameter $\beta$. We also found that the energy spectrum of the filtered velocity field, namely the resolved kinetic energy of the NS-$\alpha\beta$ model, contains more energy at the higher wavenumbers than the resolved kinetic energy of the NS-$\alpha$ model and, thus, provides a better approximation to  high resolution DNS results. The energy spectra of the unfiltered velocity fields of NS-$\alpha$ and NS-$\alpha\beta$ model were both found to significantly overpredict the energy at intermediate and high wavenumbers. 
Consistent with the nature of its viscous term~\eref{eq:dissFourier}, the NS-$\alpha\beta*$ model was found to produce higher energies at larger wavenumbers when compared to $256^3$ DNS results. We learned that choosing $\beta$ in accord with $\beta < \alpha$ broadens the ``raw" PDFs of the filtered and unfiltered velocity increments of the NS-$\alpha\beta$ model, leading to higher probabilities of larger velocity increments. 
This corresponds to more small scale activity with $\beta < \alpha$ and confirms the view that $\beta$ is responsible for the small scales. However, while the PDFs of the filtered velocity increments of the NS-$\alpha\beta$ model are slightly closer to the DNS results than the filtered velocity increments of the NS-$\alpha$ model, the PDFs of the unfiltered velocity increments of the NS-$\alpha\beta$ model were not found to provide good approximations to the DNS results. The structure functions of the unfiltered velocity field of the NS-$\alpha\beta$ model were seen to significantly overpredict the moments of the unfiltered velocity field, when compared to the structure functions of the DNS results. However, the structure functions of the filtered velocity field of the NS-$\alpha\beta$ model were found to exhibit better fidelity to the DNS results than the structure functions of the filtered velocity field of the NS-$\alpha$ model. For the mixed structure functions, the NS-$\alpha$ model showed better approximations to DNS results than did the NS-$\alpha\beta$ model. As opposed to the structure functions, we found that the flatness, as a nondimensional, relative statistic of the unfiltered velocity field of the NS-$\alpha\beta$ model, shows excellent agreement with the unfiltered $256^3$ DNS results over a wide range of separation distances, whereas the flatness of the filtered velocity field exhibits significantly lower values than the DNS results at small separation distances. These findings suggest that the unfiltered velocity field has intermittency properties very similar to the DNS results at higher resolutions, whereas the filtered velocity field is less intermittent. This is consistent with the idea of filtering.

Our statistical studies indicate that, in the NS-$\alpha\beta$ model, choosing $\beta$ such that $\beta < \alpha$ allows for more small scale activity in both the filtered and unfiltered velocity fields and, thus, provides good approximations to DNS results at higher resolution. Judging from the real space statistics of the NS-$\alpha\beta$ model, we deduced that the filtered velocity field has better, physically more reasonable statistical properties than does the unfiltered velocity field. The parameters $\alpha$ and $\beta$ should therefore be adjusted in a way to aim for the best possible approximation of the filtered velocity field to DNS results at higher resolution.

\ack{
DFH acknowledges the support of the Antje Graupe Pryor Foundation. EF acknowledges the support of the US Department of Energy and the Canada Research Chairs program.
}


\section*{References}

\begin{thebibliography}{99}




\bibitem{Holm2005}
Holm D D, 2005
{\it Los Alamos Science}  {\bf{29}} 172

\bibitem{Chen1998}
Chen S, Foias C,  Holm D D, Olson E, Titi E S, and
  Wynne S 1998
{\it Phys.\ Rev.\ Lett.\ } {\bf{81}} 5338 

\bibitem{Chen1999}
Chen S, Foias C, Holm D D, Olson E, Titi E S, and Wynne S 1999
{\it Physica D } {\bf{133}} 49 

\bibitem{Chen1999a}
Chen S, Foias C, Holm D D, Olson E, Titi E S, and Wynne S 1999
{\it Phys. Fluids }  {\bf{11}} 2343 

\bibitem{Holm1998a}
Holm D D, Marsden J E, and Ratiu T S 1998
{\it Adv.\ Math. } {\bf{137}} 1 

\bibitem{Holm1998}
Holm D D, Marsden J E, and Ratiu T S 1998
{\it Phys.\ Rev.\ Lett.\ } {\bf{80}} 4173 

\bibitem{Holm1999a}
Holm D D 1999
{\it Physica D } {\bf{133}} 215 

\bibitem{Chen1999b}
Chen S, Holm D D, Margolin L G, and Zhang R 1999
{\it Physica D } {\bf{133}} 66 

\bibitem{Fried2007}
Fried E and Gurtin M E 2007
{\it Phys.\ Rev.\ E }  {\bf{75}} 056306 

\bibitem{Fried2010}
Fried E and Gurtin M E 2010
{\it Phys.\ Rev.\ E } {\bf{82}} 029905 

\bibitem{Fried2008}
Fried E and Gurtin M E 2008
{\it Theor.\ Comput.\ Fluid Dyn.\ } {\bf{22}} 433 


\bibitem{Geurts2002}
Geurts B J and Holm D D 2002
{\it Turbulent Flow Computation}, ed Drikakis D and Geurts B J (London: Kluwer) p 237


\bibitem{Foias2001a}
Foias C, Holm D D, and Titi E S 2001
{\it Physica D } {\bf{152--153}} 505 


\bibitem{Foias2002}
Foias C, Holm D D, and Titi E S 2002
{\it J.\ Dyn.\ and Diff.\ Eqns.\ } {\bf{14}} 1 


\bibitem{Chen2008}
Chen X and Fried E 2008
{\it Phys.\ Rev.\ E } {\bf{78}} 046317 

\bibitem{Kim2009}
Kim T-Y, Cassiani M, Albertson J D, Dolbow J E, Fried E, and Gurtin M E 2009
{\it Phys. Rev. E } {\bf{79}} 045307 

\bibitem{Kim2011a}
Kim T-Y, Dolbow J E, and Fried E 2011
{\it Comp. Fluids } {\bf{44}}  102 

\bibitem{Kim2011}
Kim T-Y, Neda M, Rebholz L G, and Fried E 2011
{\it Comp.\ Meth.\ Appl.\ M.\ } {\bf{200}} 2891 


\bibitem{Kim2012}
Kim T-Y, Rebholz L G, and Fried E 2012
{\it J.\ Comput.\ Phys.\ } {\bf{231}} 4015 


\bibitem{Chen1993}
Chen S, Doolen G D, Kraichnan R H, and She Z-S 1993
{\it Phys. Fluids A } {\bf{5}} 458 

\bibitem{Meneveau2000}
Meneveau C and Katz J 2000
{\it Ann.\ Rev.\ Fluid Mech.\ } {\bf{32}} 1 

\bibitem{Mohseni2003}
Mohseni K, Kosovi\'{c} B, Shkoller S, and Marsden J E 2003
{\it Phys.\ Fluids } {\bf{15}}  524 

\bibitem{PietarilaGraham2007}
Pietarila Graham J, Holm D D, Mininni P D, and Pouquet A 2007
{\it Phys.\ Rev. E } {\bf{76}} 056310 

\bibitem{PietarilaGraham2008}
Pietarila Graham J, Holm D D, Mininni P D, and Pouquet A 2008
{\it Phys.\ Fluids } {\bf{20}} 035107 

\bibitem{Geurts2008}
Geurts B J, Kuczaj A K, and Titi E S 2008
{\it J.\ Phys.\ A } {\bf{41}} 344008 

\bibitem{Ishihara2009}
Ishihara T, Gotoh T, and Kaneda Y 2009
{\it Ann.\ Rev.\ Fluid Mech.\ } {\bf{41}} 165 

\bibitem{Frisch1996}
Frisch U 1996
{\em {Turbulence: The Legacy of A.\ N.\ Kolmogorov}}
(Cambridge: Cambridge Univ.\ Press)

\bibitem{Cao1996}
Cao N, Chen S, and She Z-S 1996
{\it Phys.\ Rev.\ Lett.\ } {\bf{76}}  3711 

\end{thebibliography}
\end{document}